\documentclass[usenatbib]{mn2e}

\usepackage{graphicx}
\usepackage{graphics}
\usepackage{deluxetable}
\usepackage{color}
\usepackage{url} 
\usepackage{hyperref}
\usepackage{booktabs}
\usepackage{float}
\usepackage{longtable}
\usepackage{color}
\usepackage{comment}

\def\gtrsim{\mathrel{\hbox{\rlap{\hbox{\lower5pt\hbox{$\sim$}}}\hbox{$>$}}}}
\def\lesssim{\mathrel{\hbox{\rlap{\hbox{\lower5pt\hbox{$\sim$}}}\hbox{$<$}}}}

\def\actaa{\rm Acta. Astr.}
\def\apj{\rm ApJ}
\def\apjl{\rm ApJL}
\def\aj{\rm AJ}
\def\mnras{\rm MNRAS}

\def\pasp{\rm PASP}
\def\aap{\rm AAP}
\def\araa{\rm ARA\&A}
\def\nat{\rm Nature}

\def\ssr{\rm Space Sci. Rev.}

\title[Observations of Supernova 2010jl]{Optical observations of the luminous Type IIn Supernova 2010jl for over 900 days}
\author[J.~E. Jencson et~al.]{J.~E. Jencson$^{1,7}$\thanks{E-mail: jj@astro.caltech.edu}, J.~L. Prieto$^{2,3}$, C.~S. Kochanek$^{4,5}$, B.~J. Shappee$^{6,8,9}$, K.~Z. Stanek$^{4,5}$
\newauthor
and R.~W. Pogge$^{4,5}$\\
$^{1}$Cahill Center for Astronomy and Astrophysics, California Institute of Technology, Pasadena, CA 91125, USA\\
$^{2}$N\'ucleo de Astronom\'ia de la Facultad de Ingenier\'ia, Universidad Diego Portales, Av. Ej\'ercito 441, Santiago, Chile\\
$^{3}$Millennium Institute of Astrophysics, Santiago, Chile\\
$^{4}$Department of Astronomy, The Ohio State University, 140 West 18th Avenue, Columbus, OH 43210, USA\\
$^{5}$Center for Cosmology and AstroParticle Physics (CCAPP), The Ohio State University, 191 West Woodruff Avenue, Columbus, OH 43210\\
$^{6}$Carnegie Observatories, 813 Santa Barbara Street, Pasadena, CA 91101, USA\\
$^{7}$NSF Graduate Fellow\\
$^{8}$Hubble Fellow\\
$^{9}$Carnegie--Princeton Fellow}
\begin{document}

\maketitle

\begin{abstract}
The luminous Type IIn Supernova~(SN)~2010jl shows strong evidence for the interaction of the SN ejecta with dense circumstellar material (CSM). We present observations of SN~2010jl for $t \sim 900~\mathrm{d}$ after its earliest detection, including a sequence of optical spectra ranging from $t = 55$ to $909~\mathrm{d}$. We also supplement our late time spectra and the photometric measurements in the literature with an additional epoch of new, late time $BVRI$ photometry. Combining available photometric and spectroscopic data, we derive a semi-bolometric optical light curve and calculate a total radiated energy in the optical for SN~2010jl of $\sim 3.5\times10^{50}~\mathrm{erg}$. We also examine the evolution of the H$\alpha$ emission line profile in detail and find evidence for asymmetry in the profile for $t \gtrsim 775~\mathrm{d}$ that is not easily explained by any of the proposed scenarios for this fascinating event. Finally, we discuss the interpretations from the literature of the optical and near-infrared light curves, and propose that the most likely explanation of their evolution is the formation of new dust in the dense, pre-existing CSM wind after $\sim 300~\mathrm{d}$.
\end{abstract}


\section{Introduction}\label{sec:intro}

Type IIn supernovae (SNe~IIn) are a rare class of events, making up only $6 - 9~\mathrm{per~cent}$ of core collapse SNe \citep[e.g.,][]{smartt09, li11, smith11a}. They are characterized by the presence of strong, composite-profile emission lines of H and He. Generally, the emission line profiles of SNe~IIn may consist of narrow-width (NW, $\sim 100~\mathrm{km~s}^{-1}$) and intermediate-width (IW, $\sim 1000~\mathrm{km~s}^{-1}$) components \citep{schlegel90, filippenko97}. The NW component is often explained as arising in dense, slowly expanding circumstellar material (CSM) photoionized by the initial flash of the supernova (SN) explosion \citep[e.g.,][]{chevalier94, salamanca98, chugai02, salamanca02}. IW lines arise at early times as either Thomson scatter broadening of the NW lines or emission from cool, accelerated, post-shock CSM gas \citep[e.g.,][]{chugai94, chugai01, chugai04, dessart09, smith10}. The shock interaction of the ejecta with the dense CSM also provides a possible explanation for the extreme optical luminosity of many SNe~IIn \citep{chugai94}.

SN 2010jl was discovered in the galaxy UGC~5189A on 2010 November 3.52 (UT) by \citet{newton10}. It was classified as a SN IIn due to the presence of NW emission lines in a spectrum taken on 2010 November 5 \citep{benetti10}. \citet{smith11b} identified a possible luminous blue progenitor star in archival \textit{Hubble Space Telescope} (\textit{HST}) WFPC2 data, but it is not possible to distinguish between a single massive star and a star cluster from these observations. Observations of this region after the SN light fades could be used to differentiate between these possibilities. Regardless, each of the progenitor scenarios explored by \citet{smith11b} point to a progenitor with a mass of at least $30~M_\odot$. Spectropolarimetry results from SN~2010jl two weeks after its discovery indicate possible asymmetry in the explosion geometry, or, more likely, in the CSM \citep{patat11}. 

Pre-discovery images from the All-Sky Automated Survey (ASAS) North telescope in Hawaii \citep{pojmanski02, pigulski09} also show that SN~2010jl is intrinsically luminous, reaching a peak absolute $I$-band magnitude of $\sim -20.5$ \citep{stoll11}, placing it near the class of $\leq -21~\mathrm{mag}$ `superluminous SNe' defined by \citet{galyam12}. Additionally, \citet{stoll11} show that the host of SN~2010jl is low metallicity, supporting an emerging trend that optically luminous SNe tend to go off in low-metallicity, low-luminosity hosts \citep[e.g.,][]{kozlowski10, neill11, li11}. The ASAS images indicate that the explosion likely occurred in early October 2010. We will refer to days $t$ since the earliest detection of SN~2010jl on 2010 October 9 by ASAS (JD = 2,455,479.14; \citealp{stoll11}) throughout this paper.

\citet{andrews11} construct a spectral energy distribution using optical, and near- and mid-infrared photometry of SN~2010jl at $t\sim 90~\mathrm{d}$, and find evidence for a $T \sim 750~\mathrm{K}$ infrared (IR) excess. They interpret the IR excess as evidence for pre-existing dust in the unshocked CSM, heated by radiation from the SN. \citet{smith12} observe a systematic blueshift for the first seven months in the IW component of H$\alpha$ that arises in the post-shock CSM, indicating the formation of new dust in the post-shock cooling zone of the SN. Their spectra also indicate that the blueshift is stronger at shorter wavelengths, which one would expect if it is the result of dust formation; however, this explanation conflicts with the low dust emission temperature observed by \citet{andrews11}. An alternative explanation is that the high optical depth continuum continues to block light from the receding side of the post-shock shell, resulting in line profiles that appear blueshifted \citep{smith12}. 

An extensive data set consisting of $UBVRI$ light curves and a sequence of optical spectra for over 500 days was presented in \citet{zhang12}. Both the slow decline of the light curves and the remarkable strength of the H$\alpha$ emission with respect to the continuum for over 500 days provide strong evidence for successive interactions of the SN ejecta with a dense, hydrogen-rich CSM. They also observe a systematic blueshift of the line profiles of H$\alpha$, but disfavor the interpretation of new dust formation due to a decrease in the blueshift of the line profiles for $t\gtrsim500~\mathrm{d}$, and the lack of a corresponding decline in the optical light curves one would expect from extinction by new dust.

\citet{maeda13} claim to find strong evidence for dust formation in SN~2010jl at $t\sim550~\mathrm{d}$. They present near-IR (NIR) spectroscopy and $JHK$ photometry, in addition to optical spectroscopy and $BVRI$ photometry at similar epochs. They detect a NIR thermal component with temperatures of $T\sim1000-2000~\mathrm{K}$ which was not present at earlier epochs. They also report an increase in the blueshift of H$\alpha$ to $\gtrsim700~\mathrm{km~s}^{-1}$ for $t\gtrsim400~\mathrm{d}$, and that the degree of the blueshift is smaller for lines at longer wavelengths. Finally, they point out that the decline in optical luminosity accelerates after $\sim 1~\mathrm{yr}$. They conclude that the cumulative evidence provides a strong case for new dust formation in a dense cooling shell formed by the interaction of the SN ejecta and the CSM. 

A more complete data set including $uBVRiJHK$ photometry, ultraviolet (UV) spectra from $HST$, extensive optical spectroscopy, and spectra in the NIR was presented in \citet{fransson14}, in which the authors debate the conclusions drawn by \citet{maeda13}. They derive a `pseudo-bolometric' (optical through NIR) light curve of SN~2010jl, and show that the NIR component, which they interpret as an echo from pre-existing CSM dust heated by the SN, becomes an increasingly important contribution to the bolometric luminosity with time. They present a sequence of H$\alpha$ profiles from $t=31$ to $t=1128~\mathrm{d}$, and claim that the broad component shows smooth, peaked, and symmetric profiles characteristic of electron scattering. The blueshift present in the lines for $t\gtrsim50~\mathrm{d}$ is interpreted as a macroscopic bulk velocity due to radiative acceleration of the pre-shock scattering medium by the energetic radiation field of the SN. From the narrow CSM lines present in their spectra, some of which show a P~Cygni absorption component, they infer a CSM velocity of $\sim 100~\mathrm{km~s}^{-1}$ consistent with a Luminous Blue Variable (LBV) star as a possible progenitor. They find the mass loss rate to account for the bolometric light curve with CSM interaction to be $\gtrsim 0.1~M_\odot~\mathrm{yr}^{-1}$, and a total mass lost of $\gtrsim 3~M_\odot$.

\citet{borish15} present a sequence of $1-2.4~\mu\mathrm{m}$ NIR spectra from $t=36$ to $565~\mathrm{d}$. They observe a progressive blueshift to $\sim 700~\mathrm{km~s}^{-1}$ of the broad hydrogen lines with time, similar to the evolution seen in the optical hydrogen lines. They also observe a broad, asymmetric component in the He\thinspace\textsc{i} $\lambda\lambda$10830 and 20587 lines, and a blue shoulder feature at $\sim 5000~\mathrm{km~s}^{-1}$, possibly associated with a high velocity, He-rich ejecta flow. Similar to the results of \citet{fransson14}, they observe a $T \sim 1900~\mathrm{K}$ blackbody feature in the continuum emission at $t=403~\mathrm{d}$, likely associated with dust emission.

\citet{gall14} observed SN~2010jl with the VLT/X-shooter spectrograph in a sequence of optical-NIR spectra covering the range $0.3-2.5~\mu\mathrm{m}$ for $t\lesssim250~\mathrm{d}$, and again on $t=876~\mathrm{d}$. They find evidence for wavelength-dependent extinction in their analysis of the IW emission lines, and interpret this as extinction by newly formed dust. They derive dust extinction curves for SN~2010jl and find evidence for large dust grains in excess of $1~\mu\mathrm{m}$ in size. They claim that the dust production site transitions from the CSM to the ejecta for $t\gtrsim500~\mathrm{d}$ based on the accelerated growth in dust mass at this time. 

SN~2010jl was detected by the $Swift$ on-board X-ray Telescope \citep[XRT,][]{burrows05} on 2010 November $5.0-5.8$ with an inferred X-ray luminosity of $(3.6 \pm 0.5) \times 10^{40}~\mathrm{erg~s}^{-1}$ \citep*{immler10}. $Chandra$ observations were triggered by \citet{chandra12a} to observe SN~2010jl on $t=59$ and $373~\mathrm{d}$, which demonstrated a high temperature of $\gtrsim 10~\mathrm{keV}$ for both epochs, and a high absorption column density of $\sim 10^{24}~\mathrm{cm}^{-2}$ at the first epoch that decreases by a factor of 3 at the second epoch, likely associated with absorption by the CSM. 

\citet{ofek14} also present X-ray observations of SN~2010jl from \textit{NuSTAR} and \textit{XMM-Newton}, in addition to multi-epoch \textit{Swift}-XRT observations, and public \textit{Chandra} observations \citep[PIs Pooley, Chandra;][]{chandra12b}. They also present the PTF $R$- and $g$-band, and \textit{Swift}-UVOT light curves. They apply a more general version of the model described by \citet{svirski12} for an SN shock interacting with an optically thick CSM, assuming spherical symmetry, and find a CSM mass in excess of $10~M_\odot$. Furthermore, they measure the temperature of the X-ray emission and infer a shock velocity of $\sim 3000~\mathrm{km~s}^{-1}$.

In this paper, we present an analysis of extensive optical observations of SN~2010jl over a period of $\sim 2.5~\mathrm{yr}$. Our data set adds significant coverage of the spectroscopic time series for SN~2010jl and we evaluate the the current physical models proposed in the literature to explain these observations.

The observations, primarily a sequence of optical spectra but including a supplementary epoch of late time $BVRI$ photometry, and data reduction procedures are presented in Section~\ref{sec:obs}. In Section~\ref{sec:analysis}, we describe our analysis of these data, including the generation of a semi-bolometric, optical light curve in Section~\ref{sec:bollc}, and a detailed study of the evolution of emission line profiles, particularly H$\alpha$, in Section~\ref{sec:specev}. Finally, we discuss our physical interpretation of these results and comment on previous interpretations in Section~\ref{sec:discussion}. We summarize our conclusions in Section~\ref{sec:summary}. 

\section{Observations}\label{sec:obs}

In this section, we present extensive optical spectroscopy of SN~2010jl and an additional epoch of late time photometry, and describe data reduction procedures. 

\subsection{Optical spectroscopy}\label{sec:specobs}

We obtained a total of twelve optical spectra of SN 2010jl using a number of ground based instruments: the Dual Imaging Spectrograph (DIS) on the 3.5-m ARC telescope at Apache Point Observatory (APO), the Wide Field Reimaging CCD Camera (WFCCD) on the du~Pont Telescope at Las Campanas Observatory (LCO), the Multi-Object Double Spectrograph \citep[MODS,][]{pogge10} on the Large Binocular Telescope (LBT) at Mt. Graham International Observatory (MGIO), and the Ohio State Multi-Object Spectrograph \citep[OSMOS,][]{stoll10, martini11} on the 2.4-m Hiltner Telescope at MDM Observatory. These observations began on $t = 55~\mathrm{d}$ and continue until $t = 909~\mathrm{d}$. These spectroscopic observations are summarized in Table~\ref{table:specobs}. Also listed in Table~\ref{table:specobs} are the two spectroscopic observations originally presented in \citet{stoll11}, which are used in the analysis discussed below in Sections \ref{sec:specev} and \ref{sec:Halpha}.

\begin{table}
\caption{Spectroscopic observations of SN~2010jl}
\label{table:specobs}
\begin{tabular}{@{}lccl}
\hline
UT Date & Phase & Range & Instrument \\
        & (d)   & (\AA) &            \\
\hline
2010 Nov. 6  & 28  & 3920-6800 & OSMOS 2.4 m Hiltner \\
2010 Nov. 12 & 34  & 3110-5880 & OSMOS 2.4 m Hiltner \\
2010 Dec. 3  & 55  & 3460-9500 & DIS 3.5 m ARC       \\
2010 Dec. 29 & 81  & 3660-9100 & WFCCD 2.5 m du~Pont \\
2011 Mar. 12 & 154 & 3460-9400 & DIS 3.5 m ARC       \\
2011 Nov. 17 & 404 & 3170-9890 & MODS LBT            \\
2012 Jan. 02 & 450 & 3460-9500 & DIS 3.5 m ARC       \\
2012 Jan. 20 & 468 & 3460-9700 & DIS 3.5 m ARC       \\
2012 Feb. 22 & 501 & 3460-9700 & DIS 3.5 m ARC       \\
2012 Mar. 4  & 512 & 3830-7020 & DIS 3.5 m ARC       \\
2012 Nov. 22 & 775 & 3920-6800 & OSMOS 2.4 m Hiltner \\
2013 Jan. 29 & 824 & 3660-9100 & WFCCD 2.5 m du~Pont \\
2013 Mar. 18 & 891 & 3460-9500 & DIS 3.5 m ARC       \\
2013 Apr. 6  & 909 & 3660-9100 & WFCCD 2.5 m du~Pont \\
\hline
\end{tabular}

\medskip
The phase is the number of days since earliest detection on JD = 2,455,479.14. The first two spectra were originally presented in \citet{stoll11}.
\end{table}

The Hiltner/OSMOS spectrum, the APO/DIS spectra, and the du~Pont/WFCCD spectra were reduced using standard tasks in \textsc{iraf}\footnote{\textsc{iraf} is distributed by the National Optical Astronomy Observatory, which is operated by the Association of Universities for Research in Astronomy (AURA) under cooperative agreement with the National Science Foundation.}. The LBT/MODS spectrum was reduced using routines for MODS dual-channel long-slit grating spectra following the `MODS Basic CCD Reduction with \textit{modsCCDRed}' manual\footnote{\textit{modsCCDRed} manual is available here:\\\url{http://www.astronomy.ohio-state.edu/}\\\url{MODS/Manuals/MODSCCDRed.pdf}}. Cosmic rays were removed from the images for all spectra using \textsc{lacosmic} \citep{vandokkum01}. Wavelength and flux calibrations were performed with comparison lamps and standard star spectra, respectively. The small wavelength zero-point shift between the comparison lamp wavelengths and the observed wavelengths in the spectra was also corrected. 

\begin{figure}
\centering
\includegraphics[width=\linewidth]{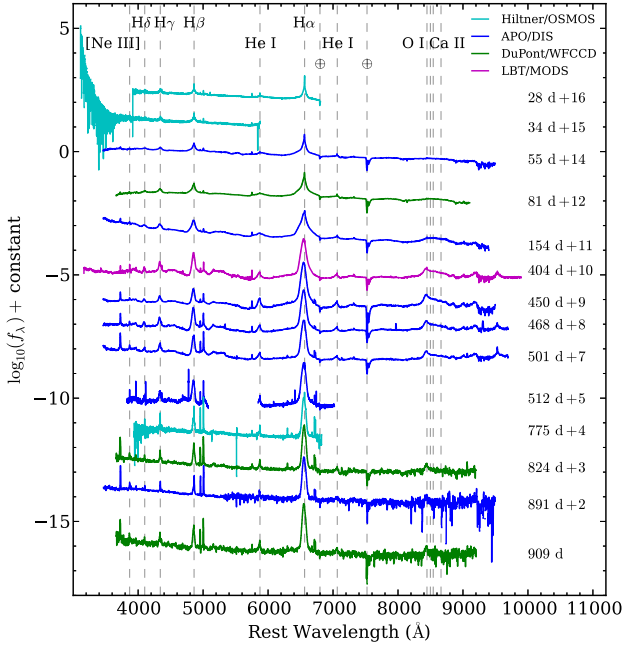}
\caption
 { \label{fig:fig1}
    Rest wavelength spectra of SN~2010jl obtained from November 2010 to April 2013. The spectra are corrected for Galactic extinction with $E(B-V) = 0.024~\mathrm{mag}$ \citep{schlafly11}. Each spectrum is shifted by the indicated constant for clarity. Days since first detection are also indicated along the right side. Also shown at the top are the two early time OSMOS spectra from \citet{stoll11}.
 }
\end{figure}

The spectra were Doppler corrected for the redshift of the host galaxy ($z=0.0107$, NED\footnote{The NASA/IPAC Extragalactic Database (NED) is operated by the Jet propulsion Laboratory, California Institute of Technology, under contract with the National Aeronautics and Space Administration.}) and for Galactic reddening using $E(B-V)=0.024$ from \citet{schlafly11}. Following \citet{smith11b}, no correction was made for host galaxy extinction. \citet{fransson14} made an estimate of the reddening for both the host galaxy and the Milky Way from the damping wings of the Ly$\alpha$ absorption in their spectra, and they adopted a value of $E(B-V)=0.058$ for the total reddening. Our spectra are shown in Fig.~\ref{fig:fig1} showing the log of the flux plus an arbitrary constant for clarity. We also show the two early time OSMOS spectra from \citet{stoll11}.

\subsection{Optical photometry}\label{sec:photobs}

\begin{figure}
\centering
\includegraphics[width=\linewidth]{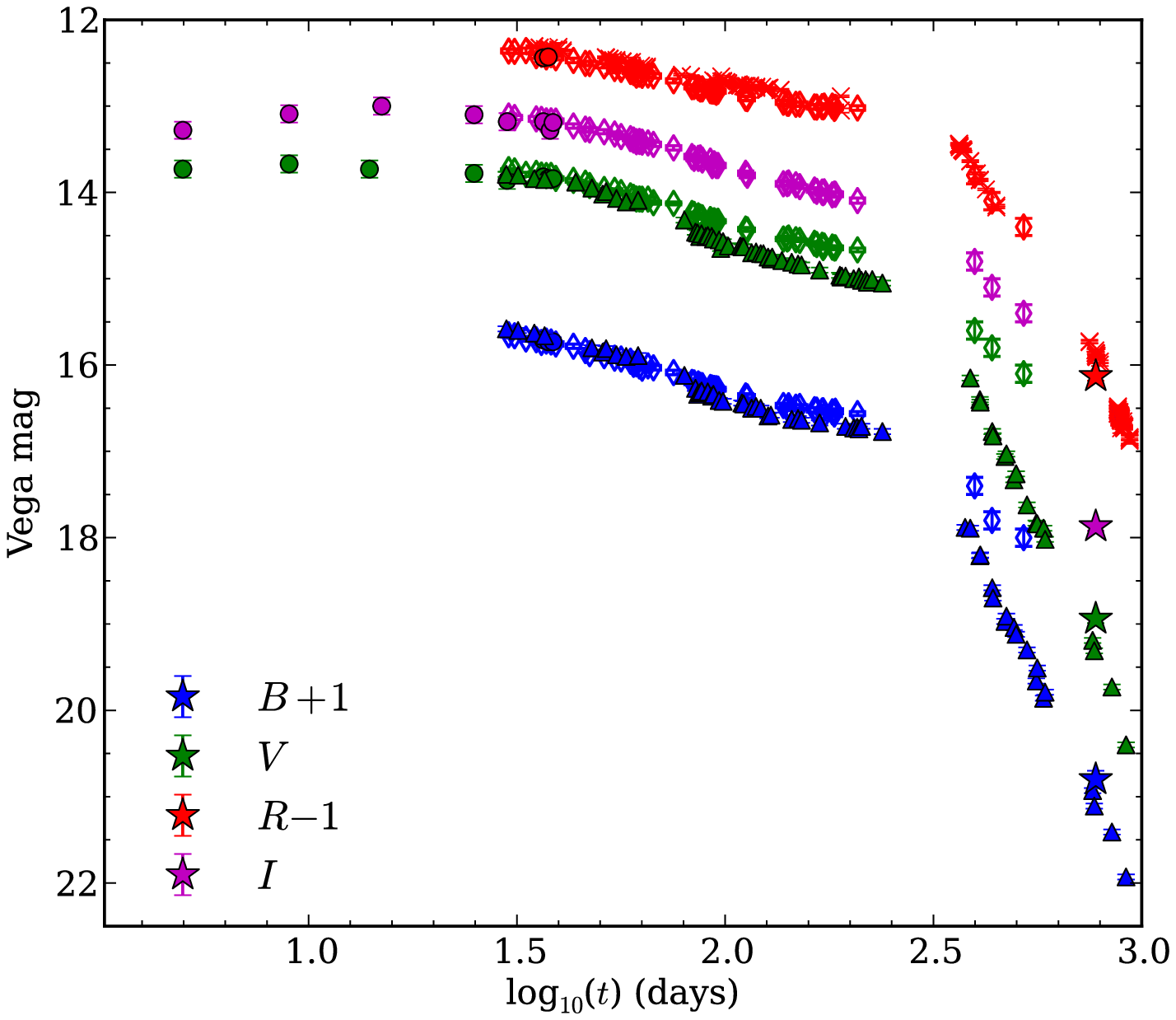}
\caption{
$BVRI$ light curves of SN~2010jl. The stars are the new data presented in this paper. Circles are early time measurements from \citet{stoll11}. The diamonds come from \citet{zhang12}. $B$- and $V$-band measurements from \citet{fransson14} are shown as triangles. The `X' symbols are the PTF $R$-band measurements from \citet{ofek14}, converted from AB to Vega magnitudes using the conversions from \citet{blanton07}. Error bars are shown, but can be smaller than the points. }
\label{fig:fig2}
\end{figure}

In order to better interpret the spectra, we performed flux calibrations (see Section~\ref{sec:bollc}) using published photometry from \citet{stoll11} and \citet{zhang12}, and obtained an additional epoch of late time $BVRI$ imaging of SN~2010jl using OSMOS on the 2.4-m Hiltner Telescope at MDM Observatory on $t=776~\mathrm{d}$. The images were reduced using standard tasks in the \textsc{iraf} \textit{ccdproc} package. Flat-fielding was performed using twilight sky flats for each filter taken from $t=773$ to $t=776~\mathrm{d}$. The photometry of the SN was obtained using the digital photometry reduction program \textsc{daophot ii} with \textsc{allstar} \citep{stetson87, stetson00}. Due to issues with over subtracting the source, we tested various annuli in \textsc{daophot} to estimate the background contribution and to minimize contamination from the host galaxy. Instrumental magnitudes were converted to apparent magnitudes using nearby field stars with SDSS DR7 \citep{dr7} $ugriz$ photometry, converted to standard $BVRI$ magnitudes using \citet{ivezic07}. Our photometric results are summarized in Table~\ref{table:phot}, where apparent magnitudes are converted to absolute magnitudes by adopting a distance modulus to the host galaxy of 33.43 mag (NED) and using the Galactic extinction corrections from \citet{schlafly11}. $BVRI$ magnitudes are also converted to band luminosities, $\nu L_{\nu}/L_\odot$, using the zero points of the Johnson-Cousins-Glass system from \citet{bessell98}. Our photometric results are shown in Fig.~\ref{fig:fig2}, along with light curves from \citet{stoll11}, \citet{zhang12}, \citet{ofek14}, and \citet{fransson14}.

\begin{table}
\caption{Photometry of SN~2010jl from $\rm{JD} = 2,456,255$ }
\label{table:phot}
\begin{tabular}{@{}cccccc}
\hline
Filter & \textit{m} & $\sigma$ & \textit{M} & $\sigma$ & $\nu L_{\nu}$ \\
       & (mag)      & (mag)    & (mag)      & (mag)    & ($L_\odot)$   \\
\hline
$B$ & $19.31$ & $0.11$ & $-14.22$ & $0.11$  & $4.3\times10^7$ \\
$V$ & $18.94$ & $0.05$ & $-14.57$ & $0.05$ & $4.1\times10^7$ \\
$R$ & $17.13$ & $0.05$ & $-16.36$ & $0.05$ & $1.6\times10^8$ \\
$I$ & $17.87$ & $0.05$ & $-15.60$ & $0.05$ & $4.9\times10^7$ \\
\hline
\end{tabular}
\end{table}

\section{Analysis}\label{sec:analysis}

In this section we present the results of our data analysis. In Section~\ref{sec:lc} we examine the new photometric data points presented in this paper in the context of published light curves from the literature. In Section~\ref{sec:bollc} we describe the procedure used to derive a semi-bolometric optical light curve for SN~2010jl, and compare its evolution to that of H$\alpha$ and the NIR luminosity. In Section~\ref{sec:specev} we describe the evolution of the optical spectra, in particular the H$\alpha$ emission profile (Section~\ref{sec:Halpha}).

\subsection{\texorpdfstring{$BVRI$}~~light curves and color evolution}\label{sec:lc}

The photometric measurements from 2012 November 23 presented in this paper are added to light curves from the literature as shown in Fig.~\ref{fig:fig2}. The light curves also include the pre-discovery $V$- and $I$-band measurements from \citet{stoll11}, the $BVRI$ light curves from \citet{zhang12}, the PTF $R$-band measurements from \citet{ofek14}, converted from AB to Vega magnitudes using the conversions by \citet{blanton07}, and the $B$- and $V$-band light curves from \citet{fransson14}. We note a few discrepancies among the measurements, particularly at late times. Beginning at $t\sim80~\mathrm{d}$, the \citet{fransson14} $B$ and $V$ light curves begin to diverge from the corresponding \citet{zhang12} light curves, with the \citet{fransson14} measurements $\sim 0.1~\mathrm{mag}$ fainter in $B$ and $\sim 0.2~\mathrm{mag}$ fainter in $V$. After the break in the light curve at $t\sim300~\mathrm{d}$, these discrepancies grow to $\sim 1.0~\mathrm{mag}$, and we note that the \citet{fransson14} light curves show a significantly steeper decline in both bands. Our measurements at $t=776~\mathrm{d}$, shown as stars in Fig.~\ref{fig:fig2}, are $\sim 0.2~\mathrm{mag}$ fainter in $R$ than the PTF measurements and $\sim 0.2 - 0.3$ mag brighter in $B$ and $V$ than the \citet{fransson14} measurements. These discrepancies may be due to difficulties in estimating the background flux as the SN fades and the contribution of the host galaxy becomes more significant. 

\begin{figure}
\centering
\includegraphics[width=\linewidth]{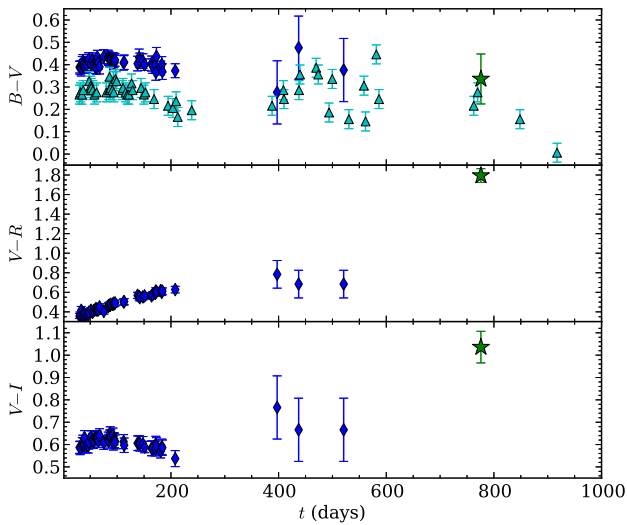}
\caption
 { \label{fig:fig3}
Color evolution of SN~2010jl. The green stars were obtained using the $BVRI$ photometry presented in this paper. The blue diamonds come from \citet{zhang12}, and the cyan triangles come from \citet{fransson14}. All measurements were corrected for Galactic reddening using the extinction corrections from \citet{schlafly11} before computing colors. Error bars are shown, but can be smaller than the points.}
\end{figure}

Fig.~\ref{fig:fig3} shows color measurements from our $BVRI$ photometry as compared to the color evolution from \citet{zhang12} and \citet{fransson14}. The $B-V$ color remains mostly constant at $\sim 0.3~\mathrm{mag}$, consistent with our measurement of $B-V = 0.3 \pm 0.1~\mathrm{mag}$ at $t=776~\mathrm{d}$. In $B-V$, we again note a $\sim 0.1~\mathrm{mag}$ discrepancy between the \citet{zhang12} and \citet{fransson14} measurements. The $B-V$ color also appears to fall off toward the blue for $t\gtrsim800~\mathrm{d}$ in the \citet{fransson14} measurements, which may be associated with the bluer continua present in the last few spectra, possibly from host galaxy contamination. In $V-R$, the color becomes redder with time, increasing from $V-R \sim 0.4~\mathrm{mag}$ at early times to our measurement of $V-R = 1.8 \pm 0.1~\mathrm{mag}$ at $t=776~\mathrm{d}$. This is consistent with the continued strength of the H$\alpha$ line (contained in the $R$-band) relative to the continuum at late times (see Sections~\ref{sec:specev} and \ref{sec:Halpha}). We also note that $V-I$ becomes redder, increasing from $V-I \sim 0.6~\mathrm{mag}$ at early times to our late time measurement of $V-I = 1.0 \pm 0.1~\mathrm{mag}$.

\subsection{Semi-bolometric optical light curve}\label{sec:bollc}

\begin{figure}
\centering
\includegraphics[width=\linewidth]{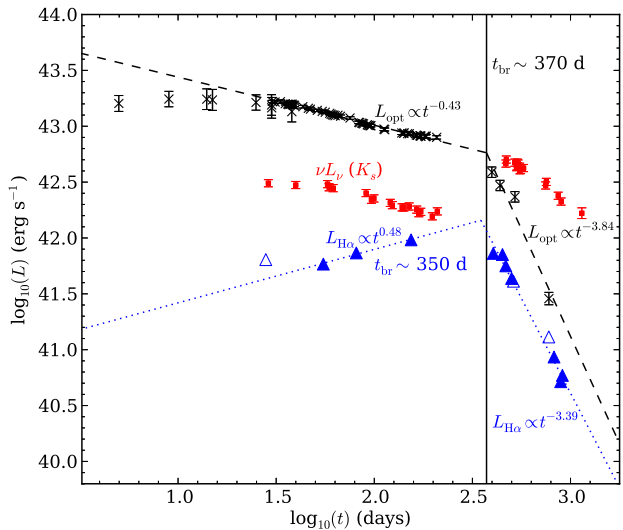}
\caption{
The semi-bolometric optical light (`X' symbols), H$\alpha$ (triangles), and $K$-band luminosities (squares) of SN~2010jl. The error bars, representing only statistical uncertainties for the derived optical points, are shown but can be smaller than the points. Power law fits to the optical light curve and H$\alpha$ are shown as black dashed and blue dotted lines. The H$\alpha$ luminosities associated with open triangles are approximate and are not included in the fits.}
\label{fig:fig4}
\end{figure}

We generated a semi-bolometric, optical light curve (3460 - 9700~\AA) using a combination of the spectra and available photometric data from this paper, \citet{stoll11} and \citet{zhang12}. We also used the 12 optical spectra from \citet{zhang12}, taken from the online Weizmann Interactive Supernova data REPository \citep[WISeREP,][]{yaron12}\footnote{WISeREP spectra are available here:\\\url{http://wiserep.weizmann.ac.il/}}. The shape of the optical spectral energy distribution (SED) was estimated as a function of time by interpolating the observed spectra, put on an absolute flux scale using synthetic photometry. Using this model of the SED with time, synthetic spectra were produced for days that had available photometry, and integrated to give the optical luminosity for each day. The resulting light curve is tabulated in Table~\ref{table:oplc} and shown in Fig.~\ref{fig:fig4}. Statistical uncertainties were estimated for each point using a bootstrap technique. The entire analysis was run 10,000 times, allowing the input photometry to vary with Gaussian distributed noise defined by the reported uncertainties. Trapezoidal integration of this light curve in time from $t=0$ to $776~\mathrm{d}$ gives an estimated total radiated energy in the optical of $(3.50\pm0.03)\times10^{50}$ erg. We caution that the uncertainty reported here is only an estimate of the statistical uncertainty from the bootstrap procedure, and does not include systematic error such as host galaxy contamination and the gap in the data when the SN was too close to the Sun to be observed between $t\sim200$ and $400~\mathrm{d}$. This number is similar to that derived by \citet{fransson14} of $\sim3.3\times10^{50}$ erg in the $3600 - 9000$~\AA\space range from $t=0$ to $920~\mathrm{d}$. In addition to the difference in wavelength and time coverage, we point out again the discrepancies between the \citet{zhang12} photometry used in this analysis and the photometric measurements used by \citet{fransson14} (see Section~\ref{sec:lc}), which may be responsible for the difference in the derived optical energy output.

Overplotted as black dashed lines in Fig.~\ref{fig:fig4} are power law fits to the derived optical light curve. The points from $t\sim20$ to $\sim370~\mathrm{d}$ were best fit with a power law decay, given by $L_{\mathrm{opt}}(t) = 1.02\times10^{43}(t/100~\mathrm{d})^{-0.43}~\mathrm{erg~s}^{-1}$, where $L_{\mathrm{opt}}$ is the optical luminosity and $t$ is the time since the earliest detection of this event. The remaining late time points through $t=776~\mathrm{d}$ were fit with an $L_{\mathrm{opt}}(t) = 6.01\times10^{42}(t/370~\mathrm{d})^{-3.84}~\mathrm{erg~s}^{-1}$ power law. The break in the light curve, defined as the intersection of the two power law fits, occurs at $t_{\mathrm{br}}\sim370~\mathrm{d}$ after first detection. The lack of data near this epoch makes it difficult to accurately constrain the timing of the break. Possible interpretations of the decline in the optical light curve at late times are discussed in Section~\ref{sec:lightcurves}.

For comparison in Fig.~\ref{fig:fig4}, we also show the H$\alpha$ luminosity as a function of time (also see Fig.~\ref{fig:fig5} and Section~\ref{sec:specev} for further discussion) and the corresponding power law fits. Luminosities corresponding to open triangles are only approximate due to insufficient wavelength coverage in the spectra for proper flux calibration with synthetic photometry and are not included in the power law fits described below. In contrast to the integrated optical luminosity, the early time H$\alpha$ luminosity follows an increasing power law given by $L_{\mathrm{H}\alpha}(t) = 7.94\times10^{41}(t/100~\mathrm{d})^{0.48}~\mathrm{erg~s}^{-1}$ until at least $t=154~\mathrm{d}$. At late times, the H$\alpha$ light curve is well fit by steep power law decay given by $L_{\mathrm{H}\alpha}(t) = 1.43\times10^{42}(t/350~\mathrm{d})^{-3.39}~\mathrm{erg~s}^{-1}$. The slope of the H$\alpha$ light curve remains slightly shallower than that of the optical light curve at late times. The break in the H$\alpha$ light curve, defined by the break in the power law fits, occurs at $t\sim 350~\mathrm{d}$, near the time of the break in the full optical light curve.

Finally, we also show the $K$-band data from \citet{fransson14} in Fig.~\ref{fig:fig4}, converted from 2MASS system magnitudes to $\nu L_{\nu}$ band luminosities using a $K_{s}$ flux zero point of $F_{\nu,0} = 666.8~\mathrm{Jy}$ at $\nu = 1.390\times10^{14}~\mathrm{Hz}$ \citep*{cohen03}. As noted by \citet{fransson14}, the NIR luminosity becomes dominant over the optical after the break in the light curve at $\sim370~\mathrm{d}$.

\begin{table}
\caption{Optical luminosity}
\label{table:oplc}
\begin{tabular}{@{}lrrr}
\hline
JD & Phase & $L$            & $L$         \\
   & (d)   & (erg s$^{-1}$) & ($L_\odot$) \\
\hline
55479.14 &   0.00 & $1.6(2)\times10^{43}$ & $4.2(4)\times10^{9}$ \\
55484.13 &   4.99 & $1.6(1)\times10^{43}$ & $4.1(3)\times10^{9}$ \\
55488.12 &   8.98 & $1.7(1)\times10^{43}$ & $4.4(3)\times10^{9}$ \\
55493.14 &  14.00 & $1.7(2)\times10^{43}$ & $4.4(4)\times10^{9}$ \\
55494.12 &  14.98 & $1.7(2)\times10^{43}$ & $4.4(4)\times10^{9}$ \\
55504.10 &  24.96 & $1.6(1)\times10^{43}$ & $4.1(3)\times10^{9}$ \\
55509.10 &  29.96 & $1.5(1)\times10^{43}$ & $3.9(4)\times10^{9}$ \\
55509.13 &  29.99 & $1.4(1)\times10^{43}$ & $3.7(3)\times10^{9}$ \\
55515.75 &  36.61 & $1.56(4)\times10^{43}$ & $4.01(9)\times10^{9}$ \\
55516.75 &  37.61 & $1.58(4)\times10^{43}$ & $4.0(1)\times10^{9}$ \\
\hline
\end{tabular}

\medskip
The phase is the number of days since earliest detection on JD = 2,455,479.14. Table~\ref{table:oplc} is published in its entirety in the electronic edition of this journal. A portion is shown here for guidance regarding its form and content.
\end{table}

\subsection{Spectral evolution of SN~2010jl}\label{sec:specev}

As seen in Fig.~\ref{fig:fig1}, the major features of the spectra of SN~2010jl remain mostly unchanged throughout its evolution, with some notable exceptions. The continuum of each spectrum can be approximated by a T~$\sim 6000-8000~\mathrm{K}$ blackbody spectrum, however, we note a slight bluing of the continuum at late times which may be due to contaminating light from the host galaxy as the SN fades. We also detect [O\thinspace\textsc{iii}] narrow emission features near 5000~\AA, which become more prominent at late times and are also likely due to contamination from the underlying star forming region. Host galaxy spectra of two H\thinspace\textsc{ii} regions $\sim 7~\mathrm{arcsec}$ north and south of the position of the SN from \citet{stoll11} clearly show strong, narrow Balmer and [O\thinspace\textsc{iii}] emission features indicative of recent star formation. A number of spectroscopic parameters, discussed below are given in Table~\ref{table:specparam}.

\begin{table*}
\begin{minipage}{140mm}
\caption{Spectroscopic parameters of SN~2010jl}
\label{table:specparam}
\begin{tabular}{@{}cccccccc}
\hline
Phase & \multicolumn{2}{c}{EW (\AA)}  & \multicolumn{2}{c}{EW (\AA)} & H$\alpha$ luminosity     & H$\alpha$/H$\beta$ & H$\beta$/H$\gamma$ \\
      & H$\alpha$ & H$\beta$ & He\thinspace\textsc{i} (5876 \AA) & He\thinspace\textsc{i} (7065 \AA) & (10$^{41}$ erg s$^{-1}$) & &       \\
\hline
28  & -167 (5)     & -35 (2)   & -14 (4)     & \nodata & 6.4  & 4.0  & 1.8 \\
34  & \nodata      & -40 (3)   & \nodata     & \nodata & \nodata & \nodata  & 1.6 \\
55  & -310 (10)    & -59 (6)   & -24 (1)     & -11 (1) & 5.8 & 5.2  & 2.5 \\
81  & -480 (20)    & -80 (6)   & -41.6 (0.7) & -18 (3) & 7.3 & 5.2  & 2.6 \\
154 & -950 (40)    & -134 (9)  & -54.6 (0.8) & -22 (2) & 9.6 & 6.1  & 3.1 \\
404 & -2700 (100)  & -300 (40) & -51 (2)     & -24 (1) & 7.3 & 6.2  & 3.6 \\
450 & -3400 (500)  & -150 (20) & -75 (2)     & -36 (2) & 7.1 & 14.3  & 3.5 \\
468 & -2400 (300)  & -270 (30) & -42 (1)     & -20 (1) & 5.6 & 6.6  & 4.0 \\
501 & -2000 (200)  & -190 (30) & -33.7 (0.9) & -19 (2) & 4.3 & 7.4  & 3.8 \\
512 & -2400 (300)  & -200 (10) & \nodata     & \nodata & 4.1  & 8.0  & 5.1 \\
775 & -1900 (200)  & -98 (5)   & -18 (2)     & \nodata & 1.3 & 10.8  & 2.9 \\
824 & -2400 (200)  & -126 (5)  & -24 (2)     & -18 (3) & 0.86 & 12.9  & 3.4 \\
891 & -2100 (100)  & -52 (2)   & -11 (2)     & -12 (2) & 0.52 & 26.1  & 3.6 \\
909 & -3000 (200)  & -150 (7)  & -34 (2)     & -14 (4) & 0.59 & 16.3  & 2.7 \\
\hline
\end{tabular}

\medskip
The phase is the number of days since earliest detection on JD = 2,455,479.14. 1-$\sigma$ uncertainties are given in parentheses. The luminosities at $t=28, 34, 512$ and $775~\mathrm{d}$ are only approximate due to insufficient spectral coverage for flux calibration with synthetic photometry. 
\end{minipage}
\end{table*}

The most prominent lines in the spectra are the strong, broad Balmer emission features. We detect the H$\alpha$ through H$\delta$ lines individually, while lines bluer than H$\delta$ are blended together. Superimposed on the broad Balmer lines, we detect narrow blueshifted P~Cygni absorption in the MODS spectrum on $t=404~\mathrm{d}$ and in the DIS spectrum on $t=512~\mathrm{d}$, but these components are not resolved in the rest of our lower resolution spectra. The width of the P~Cygni component measured from the peak to the minimum is typically $\sim100~\mathrm{km~s}^{-1}$. The evolution of the H$\alpha$ profile is discussed in detail in Section~\ref{sec:Halpha}. The broad He\thinspace\textsc{i} $\lambda\lambda$5876 and 7065 lines also show evidence for a superimposed narrow component. 

The Balmer lines, particularly H$\alpha$, show a remarkable growth in strength relative to the continuum for $t\lesssim400~\mathrm{d}$ as indicated by the equivalent width (EW) of the lines, measured using the \textsc{iraf} task \textit{splot}. From $t=28$ to $450~\mathrm{d}$, the magnitude of the EW of H$\alpha$ strengthened from $167 \pm 5$~\AA\space to $3400 \pm 500~$~\AA\space, and it remained strong at $\sim2000~$~\AA\space in each of the later spectra (See Table~\ref{table:specparam}). Similarly, the magnitude of the EW of H$\beta$ strengthened from $35 \pm 2~$~\AA\space to $300 \pm 40~$~\AA\space in approximately the first $400~\mathrm{d}$. The increase in strength of the Balmer lines is indicative of interactions of the SN ejecta with a dense hydrogen rich CSM. We also note that the He\thinspace\textsc{i} lines show similar evolution to that of the Balmer lines, but are weaker in strength.

\begin{figure}
\centering
\includegraphics[width=\linewidth]{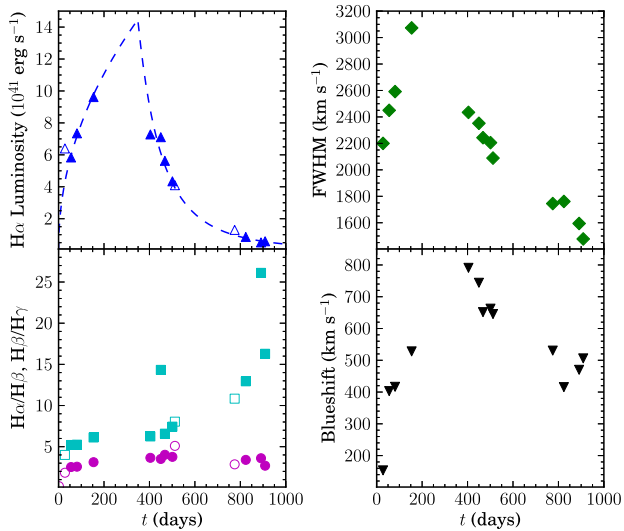}
\caption
 { \label{fig:fig5}
The evolution of various features of the Balmer emission lines. As in Fig.~\ref{fig:fig4}, open points are only approximate. Upper left: The integrated H$\alpha$ luminosity and corresponding power law fits. Lower left: The H$\alpha$/H$\beta$ and H$\beta$/H$\gamma$ luminosity ratios are shown as cyan squares and magenta dots, respectively. Upper right: The FWHM of the IW component of H$\alpha$. Lower right: The blueshift of the IW component of H$\alpha$ measured as the median velocity of the IW profile.
}
\end{figure}

As given in Table~\ref{table:specparam} and shown in the left column of Fig.~\ref{fig:fig5}, we calculate the H$\alpha$ luminosity, as well as the H$\alpha$/H$\beta$ and H$\beta$/H$\gamma$ luminosity ratios. Luminosities were calculated by subtracting a linear approximation to the continuum under the line, and then integrating the flux for the full width of the line. For $t=28, 34, 512,$ and $775~\mathrm{d}$, the luminosity measurements are only approximate due to insufficient wavelength coverage in the spectra for absolute flux calibration using available photometry. We find similar results to \citet{fransson14} (compare with their fig.~12). The H$\alpha$ luminosity increases by a factor of $\sim 1.5$ from $t=28$ to $154~\mathrm{d}$. After the break in the light curve, which occurred when the SN was close to the Sun, the luminosity drops again, and continues to fall by a factor of $\sim 10$ between $t\sim400$ and $t\sim900~\mathrm{d}$. From the power law fits shown in the figure and discussed above in Section~\ref{sec:bollc}, the maximum in the H$\alpha$ luminosity likely occurs at $t \sim 350~\mathrm{d}$. This is $\sim 10~\mathrm{d}$ earlier than the date found by \citet{fransson14}, but the lack of data between $t \sim 200$ and $t \sim 400~\mathrm{d}$ makes it difficult to put strong constraints on this date. As shown in Fig.~\ref{fig:fig5}, the H$\alpha$/H$\beta$ ratio increases from $\sim 4.0$ to $\sim 16.3$ between $t=28$ to $909~\mathrm{d}$. The H$\beta$/H$\gamma$ ratio increases slightly from $\sim 1.8$ to $\sim 5.1$ between $t=28$ and $512~\mathrm{d}$, and then decreases slightly again to around 3.0 from $t\sim800$ to $\sim900~\mathrm{d}$.

An obvious change in the appearance of the spectra is the development of blended Ca\thinspace\textsc{ii} IR triplet emission ($\lambda\lambda$8498, 8452, 8662). These lines are absent in the earliest spectra, and begin to develop by $t=81~\mathrm{d}$. The Ca\thinspace\textsc{ii} IR triplet is a more prominent feature in the spectra by $t=154~\mathrm{d}$. These emission features may be blended with a number of other lines including the O\thinspace\textsc{i} $\lambda$8446 feature. We also detect the presence of [Ne\thinspace\textsc{iii}] $\lambda$3865.9 in each of our spectra with sufficient wavelength coverage.

\subsubsection{Evolution of the \texorpdfstring{H$\alpha$}~~emission profile}\label{sec:Halpha}
 
\begin{figure}
\centering
\includegraphics[width=\linewidth]{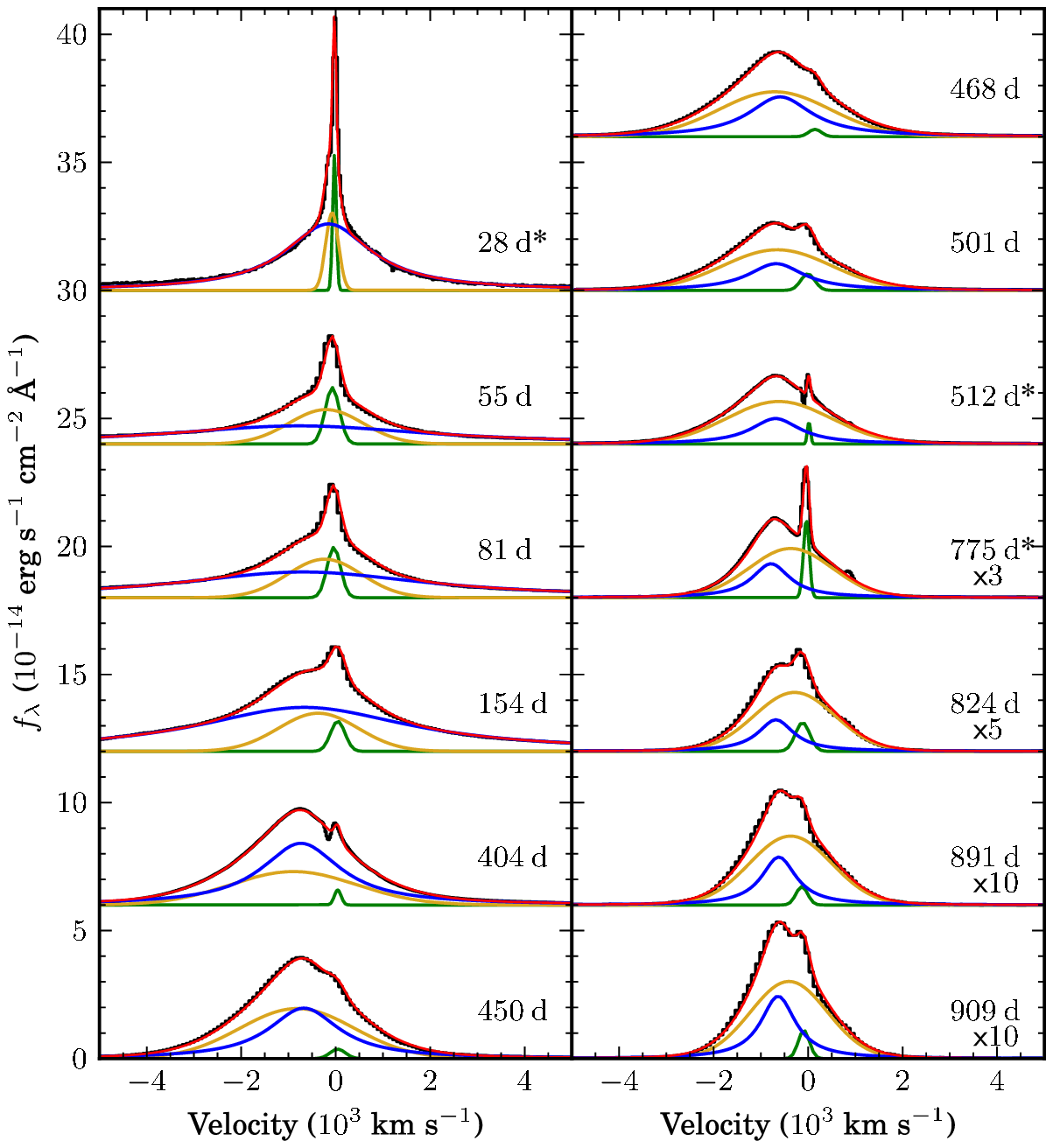}
\caption
 { \label{fig:fig6}
The continuum subtracted H$\alpha$ profiles are shown in black, where 0~km~s$^{-1}$ corresponds to the host galaxy rest frame wavelength of H$\alpha$. The components of each fit (a NW Gaussian, an IW Gaussian, and an IW Lorentzian) are shown in green, yellow, and blue, respectively. The best fit to the data is shown as the red curve. Each profile is shifted up by $6\times10^{-14}~\mathrm{erg~s}^{-1}~\mathrm{cm}^{-2}$~\AA$^{-1}$ from the one below, and the last four profiles have also been multiplied by the factor shown in the lower right corner along with phase for clarity. For the days indicated with asterisks, the fluxes are only approximate, corresponding to the open points in Fig.~\ref{fig:fig4}. 
}
\end{figure}

The evolution of the H$\alpha$ emission line is illustrated in Fig.~\ref{fig:fig6} from $t=28$ to $t=909~\mathrm{d}$. A summary of the spectroscopic parameters of the emission features, including H$\alpha$, is given in Table~\ref{table:specparam}. As discussed in Section~\ref{sec:specev}, a remarkable feature of the H$\alpha$ emission is the increase in the EW of this line until $t=450~\mathrm{d}$, and the continued strength of the line with respect to the continuum throughout the rest of the spectra, indicating strong, successive interactions of the SN ejecta with a dense hydrogen-rich CSM. 

To examine the evolution of the H$\alpha$ profile in detail, we performed multi-component fits to the profiles at each epoch using a Levenberg--Marquardt least-squares minimization algorithm. A linear approximation to the continuum near the line was subtracted from each profile before performing the fits. We find the profiles to be best fit by three separate components: an NW Gaussian, an IW Gaussian, and a broad or IW Lorentzian. The results of our fits are superimposed on the data in Fig.~\ref{fig:fig6}, and the parameters for each component, including the full IW component (sum of the IW Gaussian and IW Lorentzian), are given in Table~\ref{table:Haparam}. Previous work decomposed the H$\alpha$ profile into a two-component profile consisting of an NW component with full width at half maximum (FWHM) of $\lesssim1000~\mathrm{km~s}^{-1}$ and an IW component with FWHM velocity $\sim 2000-3500~\mathrm{km~s}^{-1}$ \citep[e.g.,][]{patat11, smith11b, smith12, zhang12, fransson14}. In this data set we find evidence for an NW and IW component, but we are unable to achieve satisfactory fits of the peaks and wings of the profiles simultaneously using only two components. Instead, we find that the IW component is itself usually best fit by two separate components, an IW Gaussian and an IW Lorentzian. The need for the extra component is illustrated further by the emerging asymmetry of the IW component at late times (see Fig.~\ref{fig:fig8} and discussion below).

\begin{table*}
\begin{minipage}{95mm}
\caption{H$\alpha$ profile component parameters}
\label{table:Haparam}
\begin{tabular}{@{}cccccc}
\hline
Phase & \multicolumn{4}{c}{H$\alpha$ component FWHM} & Shift \\
      & \multicolumn{4}{c}{(km s$^{-1}$)} & (km s$^{-1}$)    \\
	  & NW Gauss & IW Gauss & IW Lorentz & IW Full &         \\
\hline
28  & 90  & 320  & 2200 & 2200 & -153 \\
55  & 360 & 1850 & 6660 & 2450 & -403 \\
81  & 340 & 1850 & 6340 & 2590 & -417 \\
154 & 360 & 1990 & 5470 & 3070 & -528 \\
404 & 170 & 3140 & 2030 & 2430 & -791 \\
450 & 390 & 2860 & 1810 & 2350 & -744 \\
468 & 320 & 2770 & 1620 & 2240 & -651 \\
501 & 350 & 2590 & 1430 & 2210 & -663 \\
512 & 80  & 2060 & 1260 & 2100 & -646 \\
775 & 150 & 2020 & 970  & 1750 & -531 \\
824 & 360 & 2020 & 910  & 1760 & -416 \\
891 & 290 & 2040 & 890  & 1590 & -470 \\
909 & 280 & 1910 & 890  & 1480 & -506 \\
\hline
\end{tabular}

\medskip
The phase is the number of days since earliest detection on JD = 2,455,479.14. The shift is measured as the median of the full IW component.
\end{minipage}
\end{table*}

The center of the NW Gaussian component for each H$\alpha$ profile is typically consistent with $0 \pm 100~\mathrm{km~s}^{-1}$ relative to the rest frame of the host galaxy. In our highest resolution spectra, the LBT/MODS spectrum from $t=404~\mathrm{d}$ and the APO/DIS spectrum from $t=512~\mathrm{d}$, we see that the NW component is actually a P~Cygni component consisting both of emission and blueshifted absorption with a typical width of $\sim100~\mathrm{km~s}^{-1}$ measured from peak to minimum. For these profiles, the NW Gaussian is unable to fit the absorption component and is only representative of the emission component of the narrow P~Cygni profile. In the lower resolution spectra, the P~Cygni profile is unresolved, and is fit by the NW Gaussian with a typical FWHM of $\sim300~\mathrm{km~s}^{-1}$. At late times as the SN fades, the NW component may suffer from significant contamination from the underlying star forming region.

In the earliest profile from $t=28~\mathrm{d}$, the NW component of the line is best fit by both Gaussian components, one with FWHM velocity of $\sim 90~\mathrm{km~s}^{-1}$ and one with FWHM velocity of $\sim 320~\mathrm{km~s}^{-1}$. This may be due to a partially resolved, narrow P~Cygni component, resulting in the observed bump on the blue side of the narrow peak. The IW component of the profile from $t=28~\mathrm{d}$ can be approximated by a single IW Lorentzian with FWHM velocity of $\sim 2200~\mathrm{km~s}^{-1}$. 

In the subsequent profiles, the IW component, composed of the IW Gaussian and IW Lorentzian, shows a number of important changes as the profile evolves. In the profiles from $t=55, 81,$ and $154~\mathrm{d}$, the high velocity wings of the IW component are dominated by a broader (FWHM~$\sim 6000~\mathrm{km~s}^{-1}$) Lorentzian component. By $t=404~\mathrm{d}$ the high velocity wings have vanished, and the IW Lorentzian component becomes considerably narrower (FWHM $\lesssim 2000~\mathrm{km~s}^{-1}$). By $t=450~\mathrm{d}$, the full IW component begins to be more dominated by the Gaussian component (FWHM $\sim 2000-3000~\mathrm{km~s}^{-1}$). The FWHM of the IW component is plotted in the upper right panel of Fig.~\ref{fig:fig5} as a function of time. The IW component increases in FWHM to $\sim3070~\mathrm{km~s}^{-1}$ at $t=154~\mathrm{d}$, before beginning to decrease for $t\geq404~\mathrm{d}$ toward $\sim1500~\mathrm{km~s}^{-1}$ at $t=909~\mathrm{d}$. We note the qualitative similarities in the evolution of the H$\alpha$ IW component FWHM and the integrated H$\alpha$ luminosity shown in the upper left panel of Fig.~\ref{fig:fig5}. 

The profile is mostly symmetric in the Hiltner/OSMOS spectrum on $t=28~\mathrm{d}$. By $t\sim 50~\mathrm{d}$, however, the profiles show a progressive enhancement toward the blue, while the red side appears suppressed. We also note a progressive blueshift in the profiles, measured as the median velocity of the full IW component. The blueshift of the IW component of H$\alpha$ is shown in the bottom right panel of Fig.~\ref{fig:fig5}. The profile becomes significantly blueshifted with an IW component median velocity of $\sim 400~\mathrm{km~s}^{-1}$  on $t=55~\mathrm{d}$, and becomes more blueshifted with a median velocity of $\sim 800~\mathrm{km~s}^{-1}$ on $t=404~\mathrm{d}$. For $t>404~\mathrm{d}$, the IW component becomes less blueshifted again, with a median velocity of $\sim 500~\mathrm{km~s}^{-1}$ in the last profile on $t=909~\mathrm{d}$.

\begin{figure}
\centering
\includegraphics[width=\linewidth]{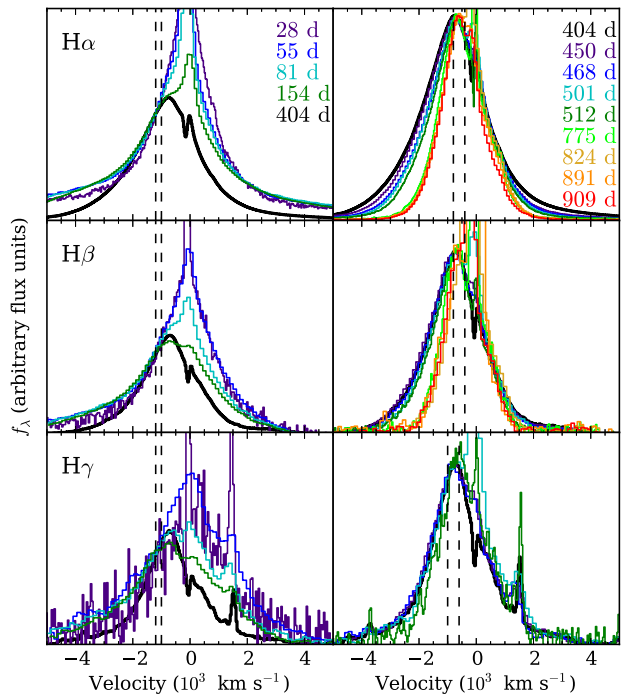}
\caption
 { \label{fig:fig7}
Sequence of emission line profiles for H$\alpha$ (upper), H$\beta$ (middle), and H$\gamma$ (lower). Early time profiles are shown in the left column, and late time profiles are shown in the right column. Profiles are scaled to match in the velocity range indicated by the dashed vertical lines in each panel. The last four H$\gamma$ profiles from $t=775$ to $909~\mathrm{d}$ were not included due to low S/N.
}
\end{figure}

\begin{figure}
\centering
\includegraphics[width=\linewidth]{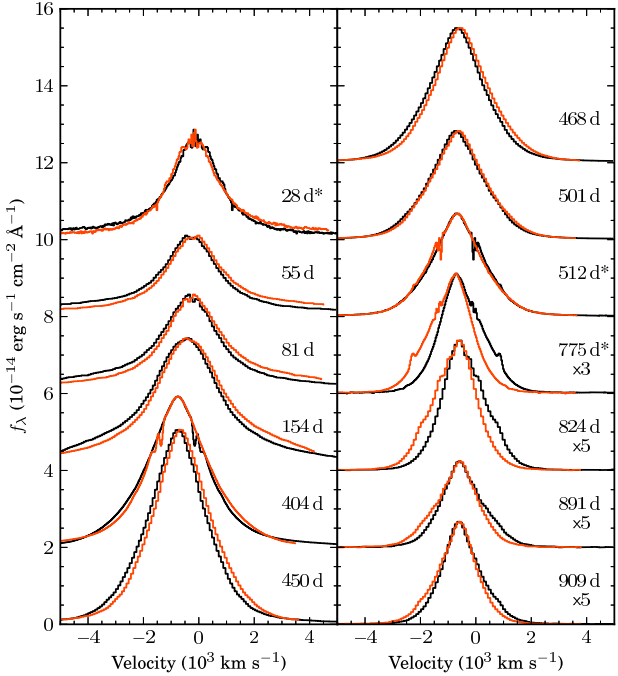}
\caption
 { \label{fig:fig8}
The continuum subtracted H$\alpha$ profiles after subtracting the NW component for clarity (black). The reflection of each profile about the IW component peak is shown in orange. Each profile is shifted up by $2\times10^{-14}~\mathrm{erg~s}^{-1}~\mathrm{cm}^{-2}$~\AA$^{-1}$ from the one below, and the last four profiles have also been multiplied by the indicated factor for clarity. Days since the earliest detection are listed for each profile. For the days indicated with asterisks, fluxes are only approximate as in Fig.~\ref{fig:fig6}. Note that the profiles appear approximately symmetric until at least $t=512~\mathrm{d}$, after which they exhibit an enhancement in flux to the red.
}
\end{figure}

The change in shape of the H$\alpha$ profile with time is illustrated in the top row of Fig.~\ref{fig:fig7}. In the top left panel, the profiles are scaled to match in the velocity range from $-1200~\mathrm{km~s}^{-1}$ to $-1000~\mathrm{km~s}^{-1}$. By $t=55~\mathrm{d}$ the profile shows signs of a relative deficit on the red side of the IW component, and this trend continues until $t=154~\mathrm{d}$. The profile from $t=404~\mathrm{d}$ is shown in black and no longer exhibits high velocity wings on either the red or blue side. A similar trend is observed for the H$\beta$ and H$\gamma$ emission line profiles, shown in the middle left and lower left panels, respectively. The late time evolution of the IW component H$\alpha$, shown in the top right panel of Fig.~\ref{fig:fig7}, changes significantly for $t\geq404~\mathrm{d}$. The profiles, now scaled to match near their peaks, continue to narrow with time, but exhibit a progressive suppression of flux toward the blue side of the peak. This results in profiles which appear asymmetric about the peak by $t=775~\mathrm{d}$.

In Fig.~\ref{fig:fig8}, we analyze the symmetric properties of the H$\alpha$ profiles in detail. The continuum subtracted profiles are shown in black as in Fig.~\ref{fig:fig6}, but the NW Gaussian component has now been subtracted for clarity. The reflection of the profile about the peak of the IW component is then shown in orange for comparison so that for a symmetric profile the orange and black curves will line up with each other. The profiles display some asymmetry from $t=28$ to $154~\mathrm{d}$, where the the flux to the red side of the peak appears suppressed, but it is possible this is due to uncertainties in estimating and subtracting the continuum. This asymmetry continues in the day 404, 450, and 468 profiles, but appears less pronounced. The profile from $t=775~\mathrm{d}$, in contrast, shows a clear suppression of flux on the blue side of the peak. This asymmetry persists in the three remaining profiles, but again becomes noticeably less pronounced with time. 

A number of scenarios have been proposed to explain the evolution of the H$\alpha$ profile, and we examine these possible interpretations below in Section~\ref{sec:IWlines}. 

\section{Discussion and physical interpretation}\label{sec:discussion}

Here we discuss the physical implications of the observed properties of SN~2010jl presented above, and provide additional commentary on physical interpretations previously presented in the literature. 

\subsection{The NW CSM lines}\label{sec:NWlines}

We observe the presence of narrow spectral emission features in the Balmer series as well as in the optical He\thinspace\textsc{i} $\lambda\lambda$5876 and 7065 lines. We interpret this as emission from slow-moving gas in the CSM, photoionized by light from the SN. In the higher resolution spectra from $t=404$ and $512~\mathrm{d}$, the narrow component of the Balmer lines exhibit blueshifted P~Cygni absorption by the expanding foreground CSM gas. The width of the P~Cygni component, measured from peak to minimum, is typically $\sim 100~\mathrm{km~s}^{-1}$, which we take as an estimate of the CSM expansion velocity. This is consistent with the estimate of the CSM velocity by \citet{fransson14}, and with the suggestion of an LBV star as a possible progenitor \citep{smith11b}. 

\subsection{The IW lines}\label{sec:IWlines}
As discussed above in Sections~\ref{sec:specev} and \ref{sec:Halpha}, the broad emission lines of H and He show a remarkable growth in strength until $t\sim450~\mathrm{d}$ and clearly indicate strong, successive interactions between the SN ejecta and dense, hydrogen-rich CSM. Similar observations of SN~2010jl have been discussed at length in the literature \citep[e.g.,][]{smith12, zhang12, fransson14, gall14}.

The IW component of the Balmer lines, especially H$\alpha$, have been of particular interest because they show some evidence for newly formed dust in the post-shock shell of the SN, but not all lines of evidence are consistent with this interpretation.

In our earliest spectrum from $t=28~\mathrm{d}$, the IW component of H$\alpha$ is well approximated by a single Lorentzian with a FWHM velocity of $\sim 2200~\mathrm{km~s}^{-1}$ (Fig.~\ref{fig:fig6}), which is characteristic of the scattering of H$\alpha$ photons by thermal electrons in an optically thick CSM. This result is consistent with the early time findings of \citet{smith12}. Soon after, however, the IW Balmer profiles begin to show significant attenuation of the redshifted flux and a progressive blueshift of the line center with time until at least $t=154~\mathrm{d}$. Based on similar findings, \citet{smith12} proposed that dust formation in the cool, post-shock shell of the SN may be obscuring red-shifted photons from interactions on the far side of the SN. \citet{gall14} observed similar attenuation of redshifted photons and progressive blueshifts of the line centers in a number of IW lines for approximately the first 240 d. They also show that this blueshift is stronger at shorter wavelengths, which would be expected if dust formation were responsible. \citet{maeda13} advocate for post-shock dust formation as the mechanism responsible for the profile shapes with similar findings at $t\sim 500~\mathrm{d}$. This interpretation is dependent on the assumption, though, that early time profiles accurately represent the intrinsic, unattenuated line profiles throughout their evolution, which my not be the case.

As is apparent in Fig.~\ref{fig:fig7}, the profile shape and evolutionary trend of the Balmer emission features has changed significantly by $t=404~\mathrm{d}$. The underlying broad component (FWHM velocity $\sim 6000~\mathrm{km~s}^{-1}$) present in earlier spectra has vanished, and the profiles continue to narrow with time throughout the remaining spectra. Rather than progressive attenuation of the red side of the profiles, we now see an increasing suppression of the flux to the blue of the IW component peak, resulting in profiles which appear very asymmetric about the peak velocity by $t=775~\mathrm{d}$ (Fig.~\ref{fig:fig8}). The profiles remain blueshifted (i.e., the majority of the line flux still falls to the blue of the host galaxy velocity) even though the degree of the blueshift appears to be decreasing for $t\lesssim 400~\mathrm{d}$. These results are apparently inconsistent with the scenario of continued dust formation at late times, since one would expect a progressive blueshift as light from the receding side of the SN is increasingly blocked by newly formed dust.

\citet{fransson14} claim that even as the H$\alpha$ profiles become more blueshifted, they retain a symmetric shape characteristic of electron scattering. They invoke electron scattering as the origin of the profile shape, with a macroscopic bulk velocity toward the observer in the scattering medium to explain the blueshift of the IW component, and reject dust formation scenarios altogether. Although the line profiles appear at least approximately symmetric about the peak velocity for the first $t\sim 500~\mathrm{d}$, the clear asymmetry present in the profiles after $t=775~\mathrm{d}$ (see Fig.~\ref{fig:fig8}) cannot be accounted for by electron scattering alone. The appearance of asymmetric and shifted line profiles could be due to asymmetries in the explosion or CSM geometries. Spectropolarimetry results from \citet{patat11} also indicated an asymmetric geometry is likely, but we do not explore this possibility further in this paper.

The formation of emission line profiles is dependent on numerous factors, including multiple velocity components and emission/absorption mechanisms. Additionally, when examining scenarios involving obscuration by newly formed, or pre-existing dust, it is impossible to know the true line profile for each epoch, and assumptions about intrinsic profile shape must be made. For these reasons, although the line profile evolution appears inconsistent with the post-shock dust formation scenario at late times, we find it difficult to rule out any scenario based solely on the shapes and evolution of emission line profiles. Further evidence from the light curves of SN~2010jl regarding the presence of newly formed or pre-existing dust are discussed below in Section~\ref{sec:lightcurves}.

\subsection{The optical and NIR light curves}\label{sec:lightcurves}
As is evident from Fig.~\ref{fig:fig4}, the optical light curve of SN~2010jl displays a clear two-stage evolution. After the peak in the optical luminosity at $t\sim 20~\mathrm{d}$, the light curve shows a steady decline which is well fit by an $L_{\mathrm{opt}}(t) = 1.02\times10^{43}(t/100~\textrm{d})^{-0.43}~\mathrm{erg~s}^{-1}$ power law decay for at least $t\lesssim200~\mathrm{d}$. At $t\sim400~\mathrm{d}$, the light curve shows a clear break and is now well fit by a much steeper power law decay given by $L_{\mathrm{opt}}(t) = 6.01\times10^{42}(t/370~\mathrm{d})^{-3.84}~\mathrm{erg~s}^{-1}$. Based on the intersection of the power law fits, the break in the light curve occurs at $t\sim 370~\mathrm{d}$. At essentially the same time, the near-IR luminosity rises rapidly so that the evolution of the bolometric luminosity changes little. A problem in the models of SN~2010jl to date is that they do not provide a natural explanation of this transition.

\citet{ofek14} lacked information on the near-IR evolution and based their model of the bolometric light curve as a simple correction to the PTF $R$-band light curve. Thus, the break in the optical light curve was interpreted as a rapid drop in the luminosity being produced by the CSM interactions. They considered three possibilities. First, the shock speed, and hence the luminosity, will drop rapidly once the swept up CSM mass begins to exceed the ejected mass (the so-called `snow plow' phase). Second, the shock could simply have reached the edge of the dense regions of the CSM. Finally, the CSM density eventually drops to the point where cooling becomes inefficient (the slow cooling stage) and a decreasing fraction of the shock energy is converted to optical photons. They rule out the last solution based on their estimate of the CSM densities, and prefer the first possibility to the second. However, with the addition of the information on the near-IR emission, there is no dramatic drop in the bolometric luminosity to be explained, or one must introduce some independent process to explain the sudden rise in the near-IR emission seen by \citet{andrews11}, \citet{maeda13}, \citet{gall14} and \citet{fransson14}. 

\citet{fransson14} interpret the two evolutionary phenomena as being unrelated. The optical luminosity drops because the shock is reaching the edge of the regions with a dense CSM or, as proposed by \citet{ofek14}, the shock slows as the swept up CSM mass becomes comparable to the ejecta mass, while the near-IR emission is caused by an echo from pre-existing dust radiatively heated by light from the SN peak. The first problem with this interpretation is that it requires remarkable coincidences in timing and energetics because the near-IR emission has to rise just as the optical emissions drops while maintaining a smooth evolution of the bolometric luminosity. The second problem, as discussed by \citet{fransson14}, is that the timing requires the inner edge of the dust shell to be at $R_{in} \sim 6 \times 10^{17}~\mathrm{cm}$, which makes it very difficult to have dust grains hot enough to produce the near-IR emission. The observed temperatures are $T_d \simeq 1500$-$2000~\mathrm{K}$ while we estimate that $a=0.1\mu\mathrm{m}$ grains would have temperatures of only $T_d=950$/$800~\mathrm{K}$ for graphite/silicate \citet{laor93} grains. \citet{fransson14} argue that this can be solved by making the grains very small, but even for a grain size of $a=10^{-3}\mu$m, we estimate dust temperatures of only $T_d=1100$/$680~\mathrm{K}$ for graphitic and silicate grains. Furthermore, for grains created in a wind, or any ejecta, there is a strong correlation between the grain size and the allowed optical depth because both are directly dependent on the density -- in essence $\tau \propto a \propto \dot{M}$. In the models of \citet{kochanek14}, a wind producing such small grains would then have a negligible optical depth at these large radii and so could not support the significant optical depth needed to produce the observed near-IR flux as an echo. Even if we are generous, the original dust-forming wind would have had at most $\tau_V \simeq 1$ and the optical depth of this large radius shell would be smaller by $R_{form}/R_{in} \simeq 0.01$ to leave an optical depth of $\tau_V \simeq 0.01$ or smaller that simply could not support such a luminous echo. 
 
The only natural way to explain a sharp break in the optical light curve combined with a simultaneous rise of the near-IR emission to an almost identical luminosity is dust forming and becoming optically thick. It not only explains the simultaneity and the similar luminosities, but also the high dust temperatures because the temperature of newly forming dust is always close to the evaporation temperature of $1500$-$2000~\mathrm{K}$. Following \citet{kochanek14} we can consider dust formation simply in terms of the `blackbody' dust temperature $\sigma T_{bb}^4 = L /16 \pi r^2$ set by the radiative flux. We will scale the results assuming dust forms at $T_{bb}=1000 T_{bb3}~\mathrm{K}$ and discuss $T_{bb}$ and the actual dust temperatures in more detail below. We will model the bolometric luminosity evolution as $L = 10^{43} L_{43} t_{300}^{-1/2}~\mathrm{erg~s}^{-1}$ where $t = 300 t_{300}~\mathrm{d}$ is scaled to the approximate time of the break in the optical luminosity evolution. 

\begin{figure}
\centering
\includegraphics[width=\linewidth]{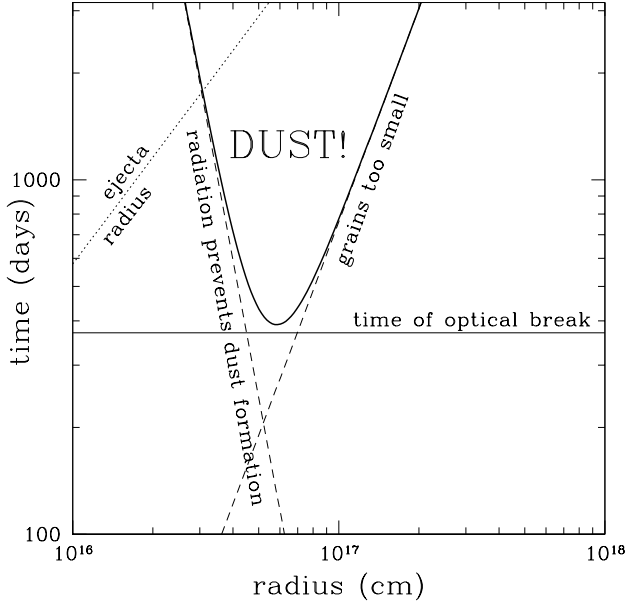}
\caption
 { \label{fig:fig9}
Scales related to dust formation. The steep, descending dashed line shows the time dependence of the radius at which the radiation will permit dust formation assuming $T_{bb3}=1$ and setting $L_{43}=0.64$ to match our estimate at $t_{300}=1$ from Section~\ref{sec:bollc}. The ejecta radius (rising dotted line) does not enter the region where dust formation is possible until long after the observed time of the optical break at 370~d (horizontal solid line). For the mass loss rate of $\dot{M}_1=0.27 v_{w2}$ invoked by \citet{ofek14}, dust grains cannot have grown large enough to have a significant visual opacity below the second dashed line. At any given radius, dust will have a significant visual opacity only at times later than the sum of these two limits, which is given by the heavy solid curve. The resulting minimum time for dust formation is very close to the observed time scale of the break in the optical light curve and the dramatic rise in the near-IR emission.
}
\end{figure}

As noted in other studies \citep[e.g.,][]{andrews11,gall14,fransson14}, dust cannot form in the material ejected during the transient because the dust temperature would be too high to form inside a radius
\begin{equation}
 R \simeq 6 \times 10^{16} L_{43}^{1/2} T_{bb3}^{-2}
    t_{300}^{-1/4}~\hbox{cm}
  \label{eqn:rdust}
\end{equation}
that is much larger than the ejecta radius 
\begin{equation}
 R \simeq 5 \times 10^{15} v_{e2} t_{300}~\hbox{cm}
\end{equation}
until long after 300 days if it is expanding at $v_e = 2000 v_{e2}~\mathrm{km~s}^{-1}$. The evolution of these two radii are shown in Fig.~\ref{fig:fig9}, and the ejecta would not reach a large enough radius to allow dust formation until $t \simeq 2000~\mathrm{d}$.

As discussed in \citet{kochanek11}, the re-obscuration of the transients SN~2008S and NGC~300 OT-2008 can be explained by dust re-condensing in their pre-existing dense winds. The CSM densities invoked to explain SN~2010jl are far denser than (even) the extreme AGB star winds invoked for these transients, although this is balanced by the far higher luminosities. Still, the optical and near-IR evolution of SN~2010jl strongly suggests that a similar process has occurred in SN~2010jl but under more extreme conditions. Inverting Equation~\ref{eqn:rdust}, the time at which dust can begin to form at a given radius is
\begin{equation}
 t_{300}^{rad} = 0.12 L_{43}^2 R_{17}^{-4} T_{bb3}^{-8}
\end{equation}
where $R = 10^{17} R_{17}~\mathrm{cm}$. This is a very rapidly falling function of radius because the luminosity is dropping so slowly. In a wind with local density $\rho_w$ where we can ignore the outward expansion on the growth time scale $t^{grow}$, a grain can collisionally grow to radius $a = f v_c \rho_w t^{grow}/4 \rho_{bulk}$ where $f=0.01 f_2$ is the condensible mass fraction, $v_c=1 v_{c1}~\mathrm{km~s}^{-1}$ is the collision velocity and $\rho_b \simeq 2.2~\mathrm{g~cm}^{-3}$ is the bulk density of the grains. If we define $\rho_w = \dot{M}/4\pi v_w R^2$ by a pre-existing wind characterized by a mass loss rate $\dot{M}= \dot{M}_1 M_\odot~\mathrm{yr}^{-1}$ and wind velocity $v_w = 100 v_{w2}~\mathrm{km~s}^{-1}$, a grain grows to radius
\begin{equation}
 a  = 0.015 { f_2 v_{c1} \dot{M}_1 t_{300}^{grow} 
         \over v_{w2} R_{17}^{-2} }~\mu\hbox{m}.
\end{equation}
Very small grains have negligible visual opacities, so we can assume there is no significant absorption until $a > a_{min} = 0.001 a_{m3}~\mu\mathrm{m}$. This then gives the required growth time at any given radius,
\begin{equation}
 t_{300}^{grow} = 0.68 { a_{m3} R_{17}^2 v_{w2} \over f_2 \dot{M}_1 v_{c1} }, 
\end{equation}
which is a rapidly rising function of radius. On small scales the radiation field keeps dust from forming and on large scales it simply grows too slowly. Fig.~\ref{fig:fig9} shows these two time scales, $t_{300}^{rad}$ and $t_{300}^{grow}$, after setting $L_{43}=0.64$ to match our estimate from Section~\ref{sec:bollc} and $\dot{M}_1=0.27 v_{w2}$ to match the wind density from \citet{ofek14}.  

Thus, there is a significant dust opacity at any given radius only once $t > t^{form}(R) = t^{rad}(R) + t^{grow}(R)$, and Fig.~\ref{fig:fig9} also shows the sum of these two time scales. Because of the different radial scalings of $t^{rad}$ and $t^{grow}$, there is a minimum time for dust formation. Moreover, without any particular manipulation of the parameters, this minimum time almost exactly matches the observed time of the optical break. If we minimize $t_{form}$, we find that the radius at which dust first reaches size $a_{min}$ is
\begin{equation}
 R_{17}^{min} \simeq 0.85 { f_2^{1/6} L_{43}^{1/3} \dot{M}_1^{1/6} v_{c1}^{1/6}
          \over a_{m3}^{1/6} T_{bb3}^{4/3} v_{w2}^{1/6} }
\end{equation}
and this occurs at time
\begin{equation}
 t_{300}^{min} \simeq 0.73 { a_{m3}^{2/3} L_{43}^{2/3} v_{w2}^{2/3}
         \over f_2^{2/3} \dot{M}_1^{2/3} T_{bb3}^{8/3} v_{c1}^{2/3} }.
\end{equation}
If we identify this time scale with the break in the light curve,
$t_b = t^{min}$, then the required mass loss rate of
\begin{equation}
 \dot{M} \simeq 0.62 { a_{m3} L_{43} v_{w2} \over f_2 v_{c1} T_{bb3}^4 t_{b300}^{3/2} }
     ~M_\odot~\mathrm{yr}^{-1}
\end{equation}
is quite close to the values of $\dot{M} \simeq 0.27 v_{w2}$ and $\simeq 0.11 v_{w2}~M_\odot~\mathrm{yr}^{-1}$ invoked by \citet{ofek14} and \citet{fransson14} to explain the observed luminosity. Once there are grains of significant size $a> a_{min}$, the optical depth becomes significant very quickly and there will always be hot, newly forming dust to support the near-IR emission.

\begin{figure}
\centering
\includegraphics[width=\linewidth]{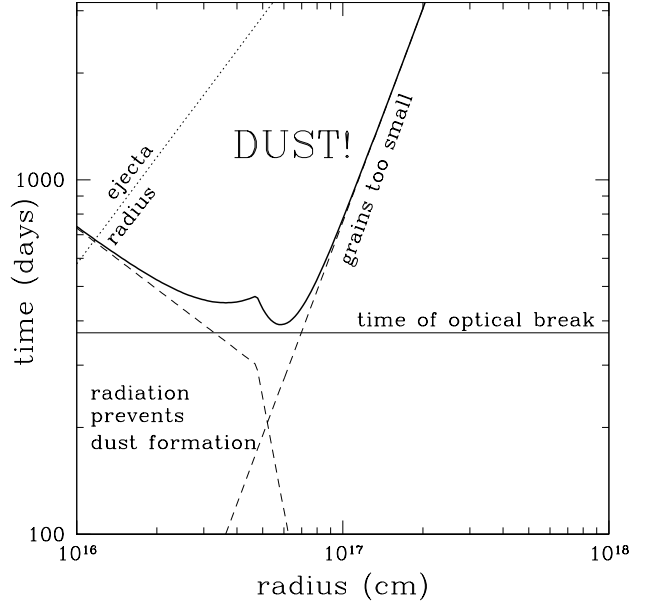}
\caption
 { \label{fig:fig10}
As in Fig.~\ref{fig:fig9} but with the total luminosity breaking from $L \propto t^{-1/2}$ to $\propto t^{-7/2}$. Shortly after the luminosity break, dust formation will occur over a broad range of radii.
}
\end{figure}

We should note that if a dense CSM exists on these radial scales, the \citet{ofek14} and \citet{fransson14} scenarios for the origin of the break make dust formation near the time of the break even more inevitable. If either the total luminosity or the radiation temperature (or both) abruptly decrease, the radius outside of which dust formation is permitted drops precipitously compared to the scenario shown in Fig.~\ref{fig:fig9}. At these smaller radii the CSM densities are higher and the grain growth times are so short that dust formation occurs almost instantaneously once permitted by the radiation field. For example, Fig.~\ref{fig:fig10} shows the effect of holding $T_{bb}$ fixed but adding a break in the luminosity from roughly $L \propto t^{-1/2}$ to $ L \propto t^{-7/2}$ we found in Section~\ref{sec:bollc}. Shortly after the luminosity break, dust formation occurs over a broad range of radii by the time $t \simeq 1000~\mathrm{d}$.

While our solution naturally explains the observations, its parameters are somewhat strained. First, given the high radiation temperature ($8000$-$9000~\mathrm{K}$) a dust formation temperature of $T_{bb} \simeq 500~\mathrm{K}$ seems more likely than the $T_{bb} = 1000~\mathrm{K}$ we have used in the scaling solutions (see \citealp{kochanek14}). But lowering $T_{bb}$ by a factor of two would require raising $\dot{M}$ by a factor of 16 to keep the $t^{min}$ fixed. Note that the dust emission would still be very hot even at $T_{bb} \simeq 500~\mathrm{K}$ because the smallest forming grains would be stochastically heated to temperatures close to their evaporation temperatures and would radiate most of their energy in these temperature spikes (see \citealp{kochanek14}). Second, once dust forms, the optical depth becomes very high, with 
\begin{equation}
  \tau_V \simeq 600 { \kappa_{V2} \dot{M}_1^{5/6} a_{m3}^{1/6} T_{bb3}^{4/3}
      \over f_2^{1/6} v_{c1}^{1/6} v_{w2}^{5/6} L_{43}^{1/3}}
\end{equation}
where we have scaled the visual opacity to $\kappa_V= 100\kappa_{V2}~\mathrm{cm}^2~\mathrm{g}^{-1}$ and this would be the optical depth for a wind extending from $R^{min}$ outwards. If it forms in a finite annulus, the optical depth is reduced by $1-R_{in}/R_{out}$ which will quickly become a modest correction (see Fig.~\ref{fig:fig9}). Empirically, the observed optical depths are far smaller.

A plausible solution is that the CSM has significant density inhomogeneities. For example, if we drop the required $T_{bb}$ to 500~K and keep $\dot{M}$ fixed while putting fraction $F=0.01$ of the CSM in clumps with an over-density of $\xi=16$, then dust will only form in the dense clumps and the mean optical depth would drop by the factor $F$. The effective drop would be still larger because of photons escaping by scattering though the low density regions. Combined with the possibilities allowed by the global asymmetries invoked by other studies of this system, this provides a means of making the properties of our solution less extreme. 

\section{Summary and conclusions}\label{sec:summary}

The luminous and long-lived Type IIn SN~2010jl shows strong evidence for the interaction of the SN ejecta with a dense, hydrogen-rich CSM shell. We obtained 12 optical spectra between $t=55$ and $909~\mathrm{d}$, which adds significant coverage to the spectroscopic time sequence for this event. We also obtained an epoch of late time $BVRI$~photometry, and derived a semi-bolometric, optical light curve through $t=775~\mathrm{d}$. 

Analysis of the spectra revealed NW ($\mathrm{FWHM}\sim100~\mathrm{km~s}^{-1}$) emission features of H and He associated with the slowing expanding CSM shell, ionized by the initial flash of the SN. The IW ($\mathrm{FWHM}\sim1000 - 2000~\mathrm{km~s}^{-1}$) component of the Balmer lines also showed complex evolution, particularly a progressive blueshift for at least the first $\sim150~\mathrm{d}$, and a pronounced asymmetry after $\sim750~\mathrm{d}$. These observations are not fully consistent with any emission and broadening mechanisms yet proposed, including post-shock dust formation and electron scattering in an optically thick medium. 

The optical light curve exhibits a steep drop off for $t\gtrsim 300~\mathrm{d}$ that has been interpreted as a transition of the SN to the snow-plow phase, or as the emergence of the shock from the CSM. However, in light of the timing coincidence of the break in the optical and rise in the NIR light curves, the most natural interpretation is that the optical light from continued CSM interactions begins to be reprocessed into the NIR after the break. Using a simple collisional model for dust growth, we find that dust naturally reforms in the dense, pre-existing CSM wind on the observed timescale of the breaks in the optical and near-IR light curves without any fine tuning of the parameters. The model would be improved if the CSM is significantly clumpy, consistent with suggestions that it has more complex structure \citep[e.g.,][]{chandra12a,maeda13,ofek14,fransson14}. We also emphasize that this interpretation indicates continued strong interactions of the SN ejecta and the CSM for $\gtrsim 2.5~\mathrm{yr}$. 

It is also interesting that a number of observational changes in the evolution of SN~2010jl are approximately coincident with the optical break and rise in the NIR light, including the decline in the H$\alpha$ luminosity, the loss of a broad, high-velocity component in the Balmer emission profiles and subsequent narrowing of the lines, and the recession in the blueshift of H$\alpha$. Despite the wealth of studies in the literature dedicated to this fascinating object, the details of the true nature of SN~2010jl's evolution remains controversial, emphasizing the importance of multiwavelength and continuous monitoring of such events.

\section*{Acknowledgments}
We would like to thank Matthew Penny and Dale Mudd for their helpful discussion and assistance in data acquisition and analysis. 

This material is based upon work supported by the National Science Foundation Graduate Research Fellowship under Grant No. DGE-1144469. BS is supported by NASA through the Hubble Fellowship grant HST-HF-51348.001 awarded by the Space Telescope Science Institute, which is operated by the Association of Universities for Research in Astronomy, Inc., for NASA, under contract NAS 5-26555. Support for JLP is in part by FONDECYT through the grant 1151445 and by the Ministry of Economy, Development, and Tourism’s Millennium Science Initiative through grant IC120009, awarded to The Millennium Institute of Astrophysics, MAS.

This work is based on observations obtained at the MDM Observatory, operated by Dartmouth College, Columbia University, Ohio State University, Ohio University, and the University of Michigan. OSMOS has been generously funded by the National Science Foundation (AST-0705170) and the Center for Cosmology and AstroParticle Physics at The Ohio State University. Based on observations obtained with the Apache Point Observatory 3.5-meter telescope, which is owned and operated by the Astrophysical Research Consortium. The LBT is an international collaboration among institutions in the United States, Italy and Germany. LBT Corporation partners are: The Ohio State University, and The Research Corporation, on behalf of The University of Notre Dame, University of Minnesota and University of Virginia; The University of Arizona on behalf of the Arizona university system; Istituto Nazionale di Astrofisica, Italy; LBT Beteiligungsgesellschaft, Germany, representing the Max-Planck Society, the Astrophysical Institute Potsdam, and Heidelberg University. This paper used data obtained with the MODS spectrographs built with funding from NSF grant AST-9987045 and the NSF Telescope System Instrumentation Program (TSIP), with additional funds from the Ohio Board of Regents and the Ohio State University Office of Research. 

Funding for the SDSS and SDSS-II has been provided by the Alfred P. Sloan Foundation, the Participating Institutions, the National Science Foundation, the U.S. Department of Energy, the National Aeronautics and Space Administration, the Japanese Monbukagakusho, the Max Planck Society, and the Higher Education Funding Council for England. The SDSS Web Site is http://www.sdss.org/. The SDSS is managed by the Astrophysical Research Consortium for the Participating Institutions. The Participating Institutions are the American Museum of Natural History, Astrophysical Institute Potsdam, University of Basel, University of Cambridge, Case Western Reserve University, University of Chicago, Drexel University, Fermilab, the Institute for Advanced Study, the Japan Participation Group, Johns Hopkins University, the Joint Institute for Nuclear Astrophysics, the Kavli Institute for Particle Astrophysics and Cosmology, the Korean Scientist Group, the Chinese Academy of Sciences (LAMOST), Los Alamos National Laboratory, the Max-Planck-Institute for Astronomy (MPIA), the Max-Planck-Institute for Astrophysics (MPA), New Mexico State University, Ohio State University, University of Pittsburgh, University of Portsmouth, Princeton University, the United States Naval Observatory, and the University of Washington.

This research has made use of the NASA/IPAC Extragalactic Database, which is operated by the Jet Propulsion Laboratory, California Institute of Technology, under contract with the National Aeronautics and Space Administration.


\begin{thebibliography}{}
\bibitem[\protect\citeauthoryear{Abazajian et al.}{2009}]{dr7} Abazajian K.~N. et al., 2009, \mnras, 351, 1071

\bibitem[\protect\citeauthoryear{Andrews et al.}{2011}]{andrews11} Andrews J.~E. et al., 2011, \aj, 142, 45

\bibitem[\protect\citeauthoryear{Benetti et al.}{2010}]{benetti10} Benetti S., Bufano F., Vinko J., Marion G.~H., Pritchard T., Wheeler J.~C., Chatzopoulos E., Shetrone M., 2010, Central Bureau Electronic Telegrams, 2536, 1

\bibitem[\protect\citeauthoryear{Bessell et al.}{1998}]{bessell98} Bessell, M.~S., Castelli, F., Plez, B., 1998, \aap, 333, 231 

\bibitem[\protect\citeauthoryear{Blanton \& Roweis}{2007}]{blanton07} Blanton M.~R., Roweis S., 2007, \aj, 133, 734

\bibitem[\protect\citeauthoryear{Borish et al.}{2015}]{borish15} Borish H.~J., Huang C., Chevalier R.~A., Breslauer B.~M., Kingery A.~M., Privon G.~C., 2015, \apj, 801, 7

\bibitem[\protect\citeauthoryear{Burrows et al.}{2005}]{burrows05} Burrows D.~N. et al., 2005, \ssr, 120, 165

\bibitem[\protect\citeauthoryear{Chandra et al.}{2012a}]{chandra12a} Chandra P., Chevalier R.~A., Irwin C.~M., Chugai N., Fransson C., Soderberg A.~M., 2012a, \apjl, 750, L2

\bibitem[\protect\citeauthoryear{Chandra et al.}{2012b}]{chandra12b} Chandra P., Chevalier R.~A., Chugai N., Fransson C., Irwin C.~M., Soderberg A.~M., Chakraborti S., Immler S., 2012b, \apj, 755, 110

\bibitem[\protect\citeauthoryear{Chevalier \& Fransson}{1994}]{chevalier94} Chevalier R.~A., Fransson C., 1994, \apj, 420, 268

\bibitem[\protect\citeauthoryear{Chugai}{2001}]{chugai01} Chugai N.~N., 2001, \mnras, 326, 1448

\bibitem[\protect\citeauthoryear{Chugai \& Danziger}{1994}]{chugai94} Chugai N.~N., Danziger I.~J., 1994, \mnras, 268, 173

\bibitem[\protect\citeauthoryear{Chugai et al.}{2002}]{chugai02} Chugai N.~N., Blinnikov S.~I., Fassia A., Lundqvist P., Meikle W.~P.~S., Sorokina E.~I., 2002, \mnras, 330, 473

\bibitem[\protect\citeauthoryear{Chugai et al.}{2004}]{chugai04} Chugai N.~N. et al., 2004, \mnras, 352, 1213

\bibitem[\protect\citeauthoryear{Cohen, Wheaton \& Megeath}{Cohen et al.}{2003}]{cohen03} Cohen M., Wheaton Wm.~A., Megeath S.~T., 2003, \aj, 126, 1090

\bibitem[\protect\citeauthoryear{Dessart et al.}{2009}]{dessart09} Dessart L., Hillier D., Gezari S., Basa S., Matheson T., 2009, \mnras, 394, 21

\bibitem[\protect\citeauthoryear{Fransson et al.}{2014}]{fransson14} Fransson C. et al., 2014, \apj, 797, 118

\bibitem[\protect\citeauthoryear{Filippenko}{1997}]{filippenko97} Filippenko A., 1997, \araa, 35, 309

\bibitem[\protect\citeauthoryear{Gall et al.}{2014}]{gall14} Gall C. et al., 2014, \nat, 511, 326

\bibitem[\protect\citeauthoryear{Gal-Yam}{2012}]{galyam12} Gal-Yam A., 2012, Science, 337, 927

\bibitem[\protect\citeauthoryear{Immler, Milne \& Pooley}{Immler et al.}{2010}]{immler10} Immler S., Milne P., Pooley D., 2010, The Astronomer's Telegram, 3012, 1

\bibitem[\protect\citeauthoryear{Ivezi\'{c} et al.}{2007}]{ivezic07} Ivezi\'{c} \u{Z}. et al., 2007, in Sterken C., ed., ASP Conf. Ser. Vol. 364, The Future of Photometric, Spectrophotometric and Polarimetric Standardization. Astron. Soc. Pac., San Francisco, p. 165

\bibitem[\protect\citeauthoryear{Kochanek}{2011}]{kochanek11} Kochanek C.~S., 2011, \apj, 741, 37

\bibitem[\protect\citeauthoryear{Kochanek}{2014}]{kochanek14} Kochanek C.~S., 2014, preprint (arxiv:1407.7856)

\bibitem[\protect\citeauthoryear{Koz{\l}owski et al.}{2010}]{kozlowski10} Koz{\l}owski S. et al., 2010, \apj, 722, 1624

\bibitem[\protect\citeauthoryear{Laor \& Draine}{1993}]{laor93} Laor A. \& Draine B.~T., 1993, \apj, 402, 441

\bibitem[\protect\citeauthoryear{Li et al.}{2011}]{li11} Li W. et al., 2011, \mnras, 412, 1441

\bibitem[\protect\citeauthoryear{Maeda et al.}{2013}]{maeda13} Maeda K. et al., 2013, \apjl, 776, 5

\bibitem[\protect\citeauthoryear{Martini et al.}{2011}]{martini11} Martini P. et al., 2011, PASP, 123, 187

\bibitem[\protect\citeauthoryear{Neill et al.}{2011}]{neill11} Neill J.~D. et al., 2011, \apj, 727, 15

\bibitem[\protect\citeauthoryear{Newton \& Puckett}{2010}]{newton10} Newton J., Puckett T., 2010, Central Bureau Electronic Telegrams, 2532, 1

\bibitem[\protect\citeauthoryear{Ofek et al.}{2014}]{ofek14} Ofek E.~O. et al., 2014, \apjl, 781, 42

\bibitem[\protect\citeauthoryear{Patat et al.}{2011}]{patat11} Patat F., Taubenberger S., Benetti S., Pastorello A., Harutyunyan A., 2011, \aap, 527, L6

\bibitem[\protect\citeauthoryear{Pigulski et al.}{2009}]{pigulski09} Pigulski A., Pojma\'{n}ski G., Pilecki B., Szczygie\l D.~M., 2009, \actaa, 59, 33

\bibitem[\protect\citeauthoryear{Pogge et al.}{2010}]{pogge10} Pogge R.~W. et al., 2010, in McLean I.~S., Ramsay S.~K., Takami H., eds, Proc. SPIE, Vol. 7735, Ground-Based and Airborne Instrumentation for Astronomy III. SPIE, Bellingham, 77350A

\bibitem[\protect\citeauthoryear{Pojma\'{n}ski}{2002}]{pojmanski02} Pojma\'{n}ski G., 2002, \actaa, 52, 397

\bibitem[\protect\citeauthoryear{Salamanca et al.}{1998}]{salamanca98} Salamanca I., Cid-Fernandes R., Tenorio-Tagle G., Telles E., Terlevich R.~J., Mu\~{n}oz-Tu\~{n}\'{o}n C., 1998, \mnras, 300, L17

\bibitem[\protect\citeauthoryear{Salamanca, Terlevich \& Tenorio-Tagle}{2002}]{salamanca02} Salamanca I., Terlevich R.~J., Tenorio-Tagle G., 2002, \mnras, 330, 844

\bibitem[\protect\citeauthoryear{Schlafly \& Finkbeiner}{2011}]{schlafly11} Schlafly E.~F., Finkbeiner D.~P., 2011, \apj, 737, 103

\bibitem[\protect\citeauthoryear{Schlegel}{1990}]{schlegel90} Schlegel E.~M., 1990, \mnras, 244, 269

\bibitem[\protect\citeauthoryear{Smartt et al.}{2009}]{smartt09} Smartt S.~J., Eldridge J.~J., Crockett R.~M., Maund J.~R., 2009, \mnras, 395, 1409

\bibitem[\protect\citeauthoryear{Smith et al.}{2010}]{smith10} Smith N., Chornock R., Silverman J.~M., Filippenko A.~V., Foley R.~J., 2010, \apj, 709, 856

\bibitem[\protect\citeauthoryear{Smith et al.}{2011a}]{smith11a} Smith N., Li W., Filippenko A.~V., Chornock R., 2011a, \mnras, 412, 1522

\bibitem[\protect\citeauthoryear{Smith et al.}{2011b}]{smith11b} Smith N. et al., 2011b, \apjl, 732, 63

\bibitem[\protect\citeauthoryear{Smith et al.}{2012}]{smith12} Smith N., Silverman J.~M., Filippenko A.~V., Cooper M.~C., Matheson T., Bian F., Weiner B.~J., Comerford J.~M., 2012, \aj, 143, 17

\bibitem[\protect\citeauthoryear{Stetson}{1987}]{stetson87} Stetson P.~B., 1987, \pasp, 99, 191

\bibitem[\protect\citeauthoryear{Stetson}{2000}]{stetson00} Stetson P.~B., 2000, User's Manual for DAOPHOT~II

\bibitem[\protect\citeauthoryear{Stoll et al.}{2010}]{stoll10} Stoll R. et al., 2010, in McLean I.~S., Ramsay S.~K., Takami H., eds, Proc. SPIE, Vol. 7735, Ground-Based and Airborne Instrumentation for Astronomy III. SPIE, Bellingham, 77354L

\bibitem[\protect\citeauthoryear{Stoll et al.}{2011}]{stoll11} Stoll R., Prieto J.~L., Stanek K.~Z., Pogge R.~W., Szczygie\l D.~M., Pojmanski G., Antognini J., Yan H., 2011, \apj, 730, 34

\bibitem[\protect\citeauthoryear{Svirski et al.}{2012}]{svirski12} Svirski G., Nakar E., Sari R., 2012, \apj, 759, 108

\bibitem[\protect\citeauthoryear{van Dokkum}{2001}]{vandokkum01} van Dokkum P.~G., 2001, \pasp, 113, 1420 

\bibitem[\protect\citeauthoryear{Yamanaka et al.}{2010}]{yamanaka10} Yamanaka M., Okushima T., Arai A., Sasada M., Sato H., 2010, Central Bureau Electronic Telegrams, 2539, 1

\bibitem[\protect\citeauthoryear{Yaron}{2012}]{yaron12} Yaron O., Gal-Yam A., 2012, \pasp, 124, 668

\bibitem[\protect\citeauthoryear{Zhang et al.}{2012}]{zhang12} Zhang T. et al., 2012, \aj, 144, 131

\end{thebibliography}
\end{document}